\documentclass{article}

\usepackage{epsfig, graphicx, amsmath}
\usepackage{dsfont}
\usepackage{amsthm}
\usepackage{xcolor}
\usepackage{bm}
\usepackage{lipsum}
\usepackage{mwe}
\DeclareMathOperator*{\argmax}{\arg\!\max}
\DeclareMathOperator*{\argmin}{\arg\!\min}

\usepackage[a4paper, total={6in, 10in}]{geometry}
\usepackage[utf8]{inputenc}
\usepackage{authblk}
\usepackage[shortlabels]{enumitem}

\usepackage{diagbox}
\usepackage[english]{babel}
\usepackage{siunitx}

\usepackage{hyperref}       
\usepackage{url}            
\usepackage{booktabs}       
\usepackage{amsfonts}       
\usepackage{nicefrac}       
\usepackage{microtype}      
\usepackage{float} 

\usepackage{algpseudocode}
\usepackage{algorithm}

\usepackage{physics}
\usepackage{subcaption}

\pdfoutput=1

\title{Programming Wireless Security through Learning-Aided Spatiotemporal Digital Coding Metamaterial Antenna}

\author{Alireza Nooraiepour$^{_{1,2}}$, Shaghayegh Vosoughitabar$^{_{1,3}}$, Chung-Tse Michael Wu$^{_{3*}}$, Waheed U. Bajwa$^{_{2}}$, Narayan B. Mandayam$^{_{2}}$}

\date{}   

\begin{document}
\maketitle

\def\thefootnote{1}\footnotetext{These authors contributed equally to this work: Alireza Nooraiepour, Shaghayegh Vosoughitabar.}\def\thefootnote{\arabic{footnote}}

\def\thefootnote{2}\footnotetext{WINLAB, Department of Electrical and Computer Engineering, Rutgers University, NJ 08902, USA.}\def\thefootnote{\arabic{footnote}}

\def\thefootnote{3}\footnotetext{Department of Electrical and Computer Engineering, Rutgers University, NJ, 08854, USA.}\def\thefootnote{\arabic{footnote}}

\def\thefootnote{*}\footnotetext{email: ctm.wu@rutgers.edu}\def\thefootnote{\arabic{footnote}}

\begin{abstract}

The advancement of future large-scale wireless networks necessitates the development of cost-effective and scalable security solutions. Conventional cryptographic methods, due to their computational and key management complexity, are unable to fulfill the low-latency and scalability requirements of these networks. Physical layer (PHY) security has been put forth as a cost-effective alternative to cryptographic mechanisms that can circumvent the need for explicit key exchange between communication devices, owing to the fact that PHY security relies on the physics of the signal transmission for providing security. In this work, a space-time-modulated digitally-coded metamaterial (MTM) leaky wave antenna (LWA) is proposed that can enable PHY security by achieving the functionalities of directional modulation (DM) using a machine learning-aided branch and bound (B\&B) optimized coding sequence. From the theoretical perspective, it is first shown that the proposed space-time MTM antenna architecture can achieve DM through both the spatial and spectral manipulation of the orthogonal frequency division multiplexing (OFDM) signal received by a user equipment. Simulation results are then provided as proof-of-principle, demonstrating the applicability of our approach for achieving DM in various communication settings. To further validate our simulation results, a prototype of the proposed architecture controlled by a field-programmable gate array (FPGA) is realized, which achieves DM via an optimized coding sequence carried out by the learning-aided branch-and-bound algorithm corresponding to the states of the MTM LWA's unit cells. Experimental results confirm the theory behind the space-time-modulated MTM LWA in achieving DM, which is observed via both the spectral harmonic patterns and bit error rate (BER) measurements.

\end{abstract}
\smallskip
\noindent \textbf{Keywords.} Physical layer security, wireless communication, metamaterial antennas, directional modulation, internet-of-things
\section{Introduction}
The last few years have witnessed a rapid proliferation of mobile users and communicating devices in the realm of wireless communication systems \cite{Ning_survey_2021,Sina_Alireza_survey2019}. In this vein, the vision of ``smart and connected" devices has been put forth by the industry, suggesting that the Internet-of-things (IoT)-driven revolution is just around the corner. For IoT networks, even the most conservative estimates are projecting tens of billions of devices by $2030$. Sustainable development of such massive networks largely depends on devising efficient low-cost security mechanisms aimed at fending off the potential malicious activities. In particular, IoT devices that have limited battery and computational resources may not be able to execute a full-blown protocol stack based on cryptographic mechanisms for security and authentication (non-access stratum) purposes \cite{Hu_2019}. As a result, increasing efforts have been made to devise a cross-layer security mechanism for securing the system at the physical layer and reducing the computational burden of encryption and authentication on the upper layers including the network layer.

Physical layer (PHY) security seeks to provide a comprehensive authentication mechanism for a communication system by exploiting the unique radio signal and electromagnetic characteristics of the communication links \cite{Sina_Alireza_survey2019,nooraiepour_TCCN,nooraiepour_MILCOM}. As a promising PHY security technique, directional modulation (DM) has enjoyed broad research attention in the last few years \cite{ding2013vector,daly2009directional}, as it has the capability of transmitting digitally modulated signals whose waveforms are well preserved only along a pre-selected direction along which legitimate user equipments are located. In practice, DM can be realized by conventional approaches for radio frequency front-ends, such as phased antenna array or digital beamforming, to provide the necessary weighting coefficients to distort (and diminish) modulated signals at undesired directions \cite{OFDM_DM,OurACM,OurIMS}. {Time-modulated arrays have also been put forth as an effective technique to achieve DM through incorporating pin diodes as radio frequency switches in the branches of a phased array to periodically connect and disconnect the antenna elements from the feeding network. However, in this approach the input signal is only transmitted in the on-time periods of the switches, leading to a decrease in signal-to-noise-ratio (SNR) at the desired angle \cite{OFDM_DM,OurIMS}.} More recently, the implementation of DM for mmWave communication has been studied through spatio-temporal transmitter arrays which are realizable in silicon chips \cite{nature_elec_1,nature_elec_2}. Nevertheless, DM phased arrays require the use of phase shifters to change phases constantly at the baseband signal modulation rate to control the excitation coefficients at the input ports  \cite{daly2009directional, daly2010demonstration}, whereas digital beamforming techniques involve bulky structures to accommodate multiple transceivers that can be very complex \cite{ding2013vector}, power consuming and expensive due to the heavy use of data converters. Moreover, both a phased array and digital beamforming approaches assume a fixed antenna array element with fixed frequency responses and radiation characteristics, in which extensive signal processing techniques are carried out separately to achieve the desired specifications for communication links, thereby increasing the cost and power consumption \cite{mendez2016hybrid}. 

Metamaterials (MTMs) consisting of sub-wavelength unit cell elements, in contrast to the conventional architectures,  have been shown to demonstrate the capability of manipulating electromagnetic waves in both transmission and reflection fashion \cite{yu2014flat,chen2016review,glybovski2016metasurfaces,Shaghayegh_IMS2020,R2-A-wireless-communication-scheme,R2-Breaking-Reciprocity}. In particular, 2D MTMs or metasurfaces can be made reconfigurable by incorporating tuning elements such as PIN or varactor diodes \cite{cui2014coding,li2017electromagnetic,zhang2018space,liu2016anisotropic}. Such programmable metasurfaces have been recently utilized as intelligent reflective surfaces \cite{R2-Breaking-Reciprocity,R2-Programmable-time-domain,R2-Realization-of-Multi-Modulation} to constantly tailor reflected electromagnetic waves in a desired fashion, which unlike the time-modulated arrays, do not decrease the SNR as they continuously reflect the incoming waveforms to their aperture. This appears to be a promising solution for next-generation communication links \cite{basar2019wireless,wu2019intelligent,wu2019towards} that can be integrated well with signal processing to adapt to the channel environments.
Programmable metasurfaces have recently been proposed in the context of backscatter communication \cite{Perfect_Absorption_Meta_Atom,MassiveBackscatter}. Specifically, \cite{Perfect_Absorption_Meta_Atom} introduces a secure backscatter communication scheme that tunes a channel environment to a special perfect-absorption condition by using programmable metasurfaces as reflectors in the radio environment. Also, a massive backscatter powered from programmable metasurfaces is proposed in \cite{MassiveBackscatter} which enables
sending beams into pre-selected directions. In addition to metasurfaces, MTMs can also be utilized to realize leaky wave antennas (LWAs) exhibiting frequency-dependent beam scanning due to their unique dispersion relationship where the propagation constant can vary from negative to positive values \cite{caloz2005electromagnetic,caloz2008crlh,jackson2012leaky}. This LWA solution is considerably low cost with respect to conventional phased arrays and can substantially reduce the design complexity and fabrication cost \cite{yuan2019multi,salarkaleji2016two}. In order to operate MTM-LWAs at a fixed frequency, tuning elements such as PIN or varactor diodes can be incorporated into the MTM unit cell to manipulate the dispersion diagram by controlling the bias voltage of tunable components \cite{lim2004metamaterial,gil2006tunable,luo2020active}. Based on this concept, very recently, dynamic metasurface antennas have been proposed for multi-input multi-output systems \cite{shlezinger2019dynamic,boyarsky2021electronically} that modulate the MTM unit cell spatially to provide required beamforming characteristics. In these metasurface antennas, only space coding has been utilized, while the time dimension has not been exploited. In other words, the coding sequences are assumed to be fixed over time and only change in accordance with the beamforming functionality requirements. Also very recently, spatiotemporally modulated metasurface antennas are proposed in \cite{R2-Sideband-free-space} that can extract and mould guided waves into any desired free-space waves in both space and frequency domains in order to overcome the issue of sideband pollution.

\begin{figure}
\centering
\includegraphics[width=0.95\textwidth, height=12cm]{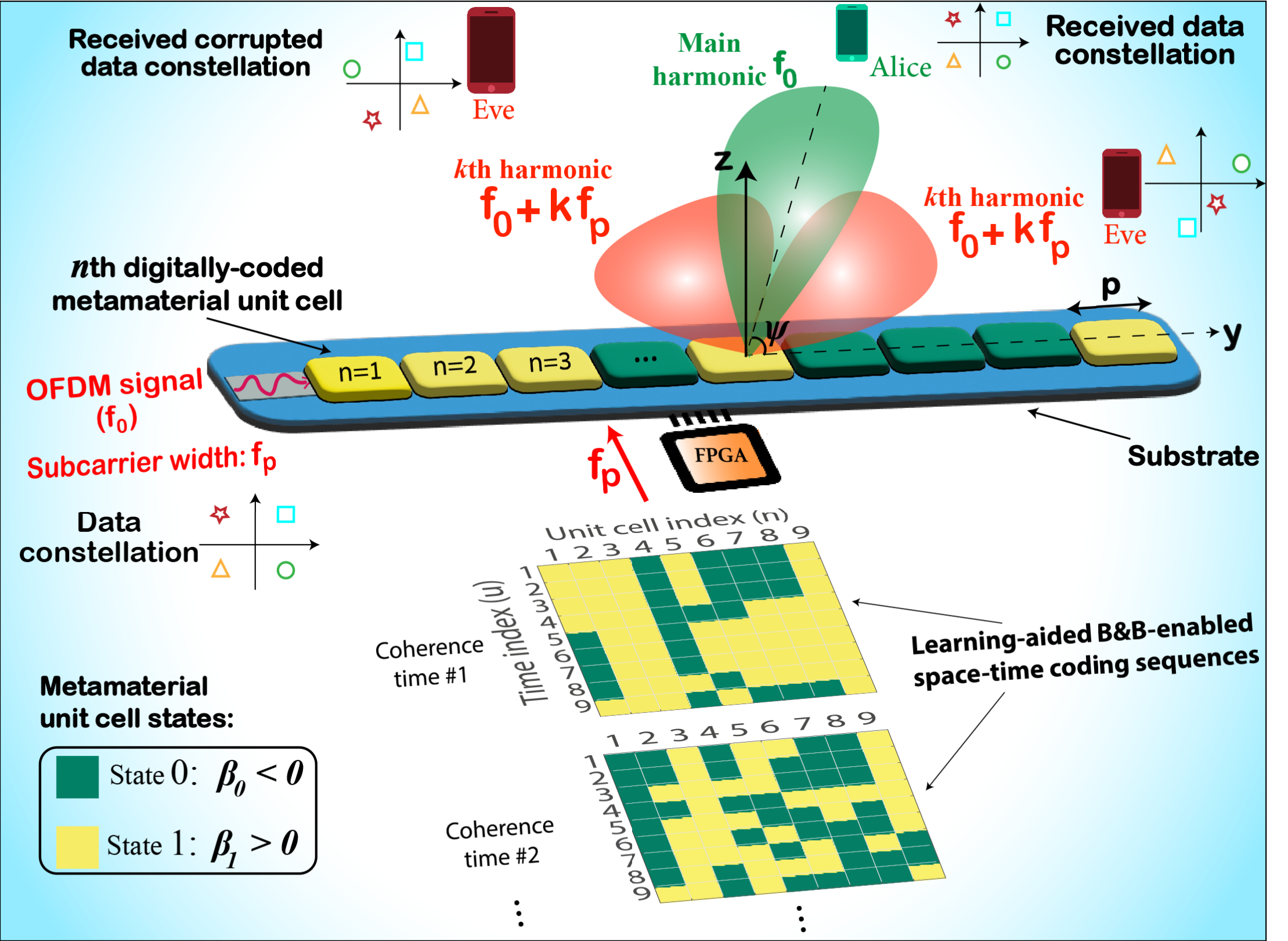}
\caption{Proposed space-time digitally-coded LWA. States of the unit cells in the LWA, either $0$ (negative phase) or $1$ (positive phase) in this case, are governed by control voltages applied to varactors in each unit cell through the FPGA, where machine learning-aided B\&B optimized coding sequence is incorporated. Time-modulation feature leads to generation and control of higher-order harmonics, while space-modulation enables beam steering for the main harmonic. In fact, for the case of free space, this architecture enables us to steer the main beam direction in the fundamental frequency ($f_0$) towards the direction of the desired legitimate receiver while suppressing the received power in the higher-order harmonic frequencies ($f_0+kf_p$) in that direction.  In this way, the original data constellation is received by the legitimate user in each OFDM subcarrier while it is distorted for the unauthorized users in all the other directions. For the case of wireless channels, similar functionality can be achieved within each coherence time by the proposed architecture as discussed later on in the paper. 
}
\label{fig:ProposedSchemes}
\end{figure}
In this work, we leverage a newly proposed space-time digitally-coded programmable MTM-LWA \cite{vosoughitabar2023programming} to achieve DM that is controlled via machine learning-aided {spatio-}temporal coding sequences, as illustrated in Figure \ref{fig:ProposedSchemes}. Specifically, in our proposed architecture, the propagation constants of the constituent composite right-/left-handed (CRLH) unit cells \cite{caloz2005electromagnetic} of a MTM-LWA are changed periodically in order to achieve the functionalities of DM, i.e., reliable communication for the desired user equipment located in a predetermined direction and security against unintended receivers in the other directions. We focus on the case where orthogonal frequency-division multiplexing (OFDM) signals, which have widely been adopted in the modern wireless communication systems including $5$G new radio, are being transmitted, leading to a multicarrier DM scheme. Our proposed scheme bridges the concepts of reconfigurable MTM antennas \cite{lim2004metamaterial,gil2006tunable,luo2020active} and time-modulated arrays \cite{ding2013vector,daly2009directional,OurACM} towards achieving DM for PHY security. Notably, our proposed transmitter architecture enables simultaneous beam steering at the fundamental frequency and beam shaping for the generated harmonics via a machine learning-aided branch and bound (B\&B) optimized coding sequence. In addition, the proposed space-time modulated CRLH unit cell can toggle between positive and negative phase constants, which will not decrease the SNR in the desired secured angle as compared to the aforementioned time-modulated arrays using pin diodes-based switches that exhibit certain off-period preventing signal transmission. While the secure backscatter communication system presented in \cite{MassiveBackscatter} utilizes space-modulated metasurfaces and a large number of meta-atoms to direct narrow beams to legitimate parties and generate noise-like signals to confuse eavesdroppers in the other directions, our proposed space-time-modulated LWA transmitter allows for the generation and manipulation of higher-order harmonics to achieve security via DM, even with a small number of antenna elements. In the remainder, we first present our theoretical derivations concerning the received signal from time-modulated MTM-LWAs in a given direction for both one and two dimensional spaces using the Fourier analysis. Secondly, we describe the underlying communication setting consisting of a pair of legitimate transmitter and receiver along with an eavesdropper. Subsequently, we formulate the mathematical problem of achieving DM as a mixed-integer non-linear program (MINLP) whose optimization variables include the coding sequence determining the states of the LWA's unit cells, and utilize the B\&B algorithm as the solver. We also take the effect of wireless channel into account in our analysis and show how the proposed architecture can achieve DM in a wireless channel setting. Due to the time-varying nature of the wireless channels, we also propose to utilize deep learning to enhance the B\&B algorithm, and obtain a learning-aided B\&B solver for finding space-time coding sequences. Finally, a space-time digitally coded MTM-LWA with $9$ unit cells, each equipped with varactors for controlling the underlying states, is fabricated and experimentally tested to verify the theoretical/simulation results.

\section{Digitally-coded space-time MTM-LWA}
\label{sec:results_theory}

We propose a space-time MTM-LWA programmed via a temporal sequence of digital codes where by applying control voltages to varactors associated with unit cells of a MTM-LWA, the resulting induced phase shift from a unit cell can be dynamically modified. As shown in Figure \ref{fig:ProposedSchemes}, these control voltages are produced by a field-programmable gate array (FPGA) and only take on quantized values that change periodically over time, resulting in a finite number of states for each unit cell at a time step. Considering  the case of binary sequences, i.e., digital codes of $0$'s and $1$'s, where the unit cells can be in two different states at a time, the time-domain
far-field radiation pattern of a digitally-coded MTM-LWA can be expressed as
\begin{align}
\label{eq:RadiationPattern1D}
R(\psi,t)=\sum_{n=1}^N S(t) e^{-\alpha(n-1)p}e^{j(n-1)k_0p\cos{\psi}}U_{n}(t),
\end{align}
where $S(t)$ is the input signal to be transmitted, $N$ is the number of unit cells, $k_0=\frac{2\pi}{\lambda}$ is the free space wavenumber, and $\lambda$ denotes the wavelength. Also, $p$, $\psi$ and $\alpha$ are the LWA's period, radiating angle, and the LWA's leakage factor, respectively. The time-modulated nature of the $n$th unit cell's contribution to the radiation pattern is attributed to a periodic phase-delay function $U_n(t)$ defined as
\begin{align}
\label{eq:U_n(t)}
U_{n}(t)=\sum_{u=1}^L \Gamma_n(\mathbf{q}_u)H^u(t),
\end{align}
over $0\leq t\leq T_p$, where $\mathbf{q}_u$ denotes the coding sequence of length $N$, $L$ denotes length of the digital codes, and $H^u(t)=\begin{cases}1,~t\in \mathcal{T}^u,\\0,~\text{else}, \end{cases}$ for $\mathcal{T}^u\overset{def}{=}[\frac{(u-1)T_p}{L},\frac{uT_p}{L}]$. Phase-delay of the $n$th unit cell during $t\in \mathcal{T}^u$ equals to $\Gamma_n(\mathbf{q}_u)= \prod_{k=1}^ne^{j\kappa_{\mathbf{q}_u(k)}}$, where $ \kappa_{\mathbf{q}_u(k)}=\begin{cases}
 \beta_0p,~\mathbf{q}_u(k)=0,\\
 \beta_1p,~\mathbf{q}_u(k)=1,
 \end{cases}$ for the case of binary sequences, and $\mathbf{q}_u(k)$ represents state of the $k$th unit cell in $\mathcal{T}^u$. We note that $\kappa_{\mathbf{q}_k(u)}$ can be modified accordingly to account for more number of states as needed. Owing to the travelling wave nature of the LWA structure, the amount of phase delay experienced by the input wave passing through the $n$th unit cell not only depends on the current state of that cell but also on the states of all the previous unit cells. As an illustrative example, when $\mathbf{q}_u=\begin{bmatrix}1,~0,~0,~1,~0,~1\end{bmatrix}$, assuming $\beta_0p=-18^\circ$ and $\beta_1p=15.5^\circ$, the phase delay incurred to the incoming wave after the sixth unit cell would be $15.5^\circ-18^\circ-18^\circ+15.5^\circ-18^\circ+15.5^\circ=-7.5^\circ$.

The periodic function $U_n(t)$ can be further expressed by using Fourier series:
 \begin{align}
 \label{eq:FS}
     U_n(t)=\sum_{\nu=-\infty}^{\infty}c_{\nu n}e^{j2\pi\nu f_pt},
 \end{align}
 where the Fourier coefficients $c_{\nu n}$ can be shown to be (see Section S$1$ of the Supporting Information)
 \begin{align}
 \label{eq:FS_coef}
     c_{\nu n}=\sum_{u=1}^L\frac{\Gamma_n(\mathbf{q}_u)}{L}\text{sinc}\big(\frac{\nu\pi}{L}\big)e^{-j\frac{\pi\nu(2u-1)}{L}}.
 \end{align}
Plugging Equations \eqref{eq:U_n(t)} and \eqref{eq:FS} in Equation \eqref{eq:RadiationPattern1D}, the radiation pattern can be simplified to $R(\psi,t)=S(t)\sum_{\nu=-\infty}^\infty w(\nu,L,t,\psi)$ where
\begin{align}
\label{eq:w_function_1D}
    w(\nu,L,t,\psi)=e^{j2\pi\nu f_pt}\frac{1}{L}\text{sinc}\big(\frac{\nu\pi}{L}\big)e^{\frac{j\pi\nu}{L}}\sum_{u=1}^L\Xi^u(\psi,\mathbf{q}_u)e^{-j\frac{2\nu\pi u}{L}},
\end{align}
and
\begin{align}
\label{eq:Xi_function_1D}
    \Xi^u(\psi,\mathbf{q}_u)=\sum_{n=1}^N\Gamma_n(\mathbf{q}_u)e^{-\alpha(n-1)p}e^{jk_0(n-1)p\cos{\psi}}.
\end{align}

The periodic changes in states of the unit cells with a period of $\frac{1}{f_p}$ cause the LWA to create an infinite number of harmonics with frequencies $\nu f_p$ and magnitudes $|w(\nu,L,t,\psi)|$ for $\nu=-\infty,\dots,\infty$ at the receiver where $\nu$ denotes the generated harmonic index. The magnitude of the harmonics at each receiving angle is a function of the number of time steps ($L$), the digital coding sequences, phase delays associated with each state ($\beta_0p$ and $\beta_1p$), and the phase shifts in each branch. We note that for the case of $L=1$ where the state of a unit cell remains fixed over time, one could steer the angle of the LWA's main beam toward a certain direction by setting the states of the unit cells. In addition to this beam-scanning feature, the time-modulated nature of the proposed LWA enables us to control the radiated spectral components.

Our proposed architecture enable the manipulation of spectral components of the received signals by the use of digital codes for any receiving angles. This is of prime applicability in multi-carrier communication systems, e.g., when the signal $S(t)$ being transmitted by the LWA is of OFDM type. For the OFDM transmission, the information is sent over a limited frequency band composed of a finite number of spectral bins called subcarriers. In order for the generated harmonics to interact with the received constellation symbols in each subcarrier and meet the DM functionalities for the case of OFDM signals, the corresponding subcarrier width is set to be the same as the switching frequency $f_p$. As a result, harmonics of the form $w(\nu,L,t,\psi)$, for $\nu\neq0$, cause interference in all the received subcarriers. The key aspect of our proposed method is to design the digital codes, i.e., $\mathbf{q}_k$ in order to control the behavior of the interference terms and enforce the way they contribute to the received signal's subcarriers in different angles. In particular, for the case of free space transmission, the digital codes would be designed in order to minimize the level of such interference for the legitimate user equipment in the desired angle, while maximizing it for the unauthorized receivers in all the other directions as highlighted in Figure \ref{fig:ProposedSchemes}. Similar functionality can also be achieved in different coherence times for the case of transmission through wireless channels as discussed in a later section, where space-time coding sequences are obtained via learning-aided B\&B algorithm.

Before delving into the specifics of designing the digital codes, we make a connection between the interference terms as a product of the time-modulated nature of transmission and a well-known notion in the physical layer security literature, i.e., secrecy capacity. This notion, which is derived based on the prominent work of Wyner \cite{Wyner}, serves as a metric to quantify both reliability and security of a communication link, and is defined as $C_s=\log(1+\text{SNR}_R)-\log(1+\text{SNR}_E)$ when $\text{SNR}_R>\text{SNR}_E$. In this definition, $\text{SNR}_R$ and $\text{SNR}_E$ refer to the SNRs at the desired and unintended UEs, respectively. Therefore, a higher SNR for the desired angles and lower SNR for the undesired angles is required to maximize the secrecy capacity. For conventional phased-array antennas, the difference between the receiving SNRs merely stems from the fact that the main beam of the radiation pattern is directed towards the desired angles while the unintended parties receive the signal through the side lobes. Although the resulting SNR from the side lobe transmission is smaller, it might still enable the unintended receivers to decode the data. The above time-modulated transmission enables one to further reduce the SNR in the undesired angles without tampering with it at the desired angle. To this end,  we design coding sequences $\mathbf{q}_u$ that generate interference at the received signal only for the undesired angles through the $w$ function in Equation \eqref{eq:w_function_1D}, which increases the noise level at the unintended receivers and reduces their corresponding SNR.

\section{DM enabled by space-time digitally-coded MTM-LWAs}
\label{sec:results_optimization}
Consider a communication setting composed of three parties: transmitter, legitimate receiver, and eavesdropper, referred to as Alice, Bob, and Eve, respectively. Bob, demonstrated as the green party in Figure \ref{fig:ProposedSchemes}, is located at a certain angle with respect to Alice. Eves, the red parties located at the other directions, are listening to the transmission, aiming to infer secret information sent by Alice to Bob. In this scenario, the goal of our proposed space-time digitally-coded MTM-LWA is to provide reliable communication to Bob, given the spatial angular information of Bob, while preventing Eves from correctly decoding the data  We assume the communication link between Alice and Bob is operated over a total bandwidth of $B$ over which Alice transmits an OFDM signal of $K$ subcarriers. Mathematically, $S(t)$ can be expressed as
$S(t)=1/K\sum_{k=1}^Ks_k e^{j2\pi(f_0+(k-1)f_p)t}$, where $f_0$, $f_p$, and $s_k$ denote the carrier frequency, subcarrier width, and complex symbol transmitted in the $k$th subcarrier, respectively. The input secret bit stream at Alice is modulated with complex symbols, also known as the constellation points (see Figure \ref{fig:ProposedSchemes}), and subsequently mapped to the subcarriers. We note that the DM technique can effectively accomplish the two aforementioned design goals of security and reliability by intentionally manipulating the spectral components received at Bob and Eve. Our proposed MTM-LWAs can fulfill the functionalities of DM \cite{OFDM_DM,OurACM} solely by the use of digital coding sequence.

In order to enable DM for Bob, the following two goals shall be achieved. First, the original transmitted constellation points corresponding to the subcarriers of the OFDM signal, $S(t)$, should be preserved along the desired angle $\psi_0$, in order to facilitate the decoding process for the receiver. As the non-zero order harmonics in the $w$ function in Equation \eqref{eq:w_function_2D} are the source of the interference introduced into the subcarriers, mathematically, we can express this constraint by $\begin{cases}w(\nu\neq0,L,t,\psi_0)=0,\\w(\nu=0,L,t,\psi_0)\neq0\end{cases}$. Second, the received constellation points at different subcarriers along all the other angles should be distorted to reduce the chance of an eavesdropper to correctly decode the transmitted data. This is equivalent to imposing that $w(\nu\neq0,L,t,\psi\neq\psi_0)\neq0$. Besides these two constraints, in order to maximize the SNR towards Bob, the angle of the main beam associated with $R(\psi,t)$, i.e., the angle at which the radiation pattern has its maximum value, should be the same as $\psi_0$ for all $t$. This amounts to $\psi_0$ being the solution of $\argmax_{\psi}|R(\psi,t)|$ for all $t$. By satisfying these three criteria, we have shown in Section S$2$ of the Supporting Information that the received signal in the desired angle is a weighted version of the transmitted OFDM signal, i.e., the complex symbols sent over the subcarriers can be reliably decoded. For the undesired angles, on the other hand, the received symbol in each subcarrier is corrupted by the interference terms.

The first two constraints mentioned above are related to the physical layer security while the third constraint pertains to the beam pattern of MTM-LWA, ensuring that the maximum amount of power is radiated in the desired direction. We have discussed in Section S$3$ of the Supporting Information why the second and the third constraints cannot be satisfied simultaneously. For the purpose of physical layer security, we relax the third constraint as $\psi_0=\argmax_{\psi}|R(\psi+d^1_u,t)|$ when $t\in\mathcal{T}_u$ by introducing slack variables $d^1_u$ for $u=1,\dots,L$, which account for the deviations of the radiation pattern's main beam angle from the desired angle at different time steps. Based on the theoretical results obtained in the previous section, we then mathematically formulate the problem of finding digital coding sequences satisfying the above three constraints for a given $\psi_0$ as (see Section S$3$ of the Supporting Information) 
\begin{align}
\label{eq:optimizationproblem_freespace}
    \argmin_{\substack{\mathbf{q}_u,d_u^1\\u=1,\dots,L}}~ \sum_{u=2}^L \big|&\Xi(\psi_0,\mathbf{q}_u)-\Xi(\psi_0,\mathbf{q}_1)\big|-\sum_{u=1}^L\big|\Xi(\psi_0+d_u^1,\mathbf{q}_u)\big|\nonumber\\&
    \text{s.t.}~~~\mathbf{q}_u\in\{0,1\}^{N},~ L_u\leq d_u^1\leq U_u,
\end{align}
where $L_u$ and $U_u$ are decimal values representing the lower and upper bounds on the deviation slack variables, respectively. As $\Xi(\psi_0,\mathbf{q}_u)$ corresponds to a LWA's radiation pattern with states $\mathbf{q}_u$, the first summation effectively computes the difference between the LWA radiation patterns for all $L$ time steps at the desired angle. In other words, the minimizer finds coding sequences whose corresponding radiation patterns have similar values for the desired angle. The second summation accounts for the fact that the maximum of the radiation patterns occurs in the vicinity of the desired angle. In fact, larger values of $d_u^1$'s translate to larger differences between the radiation patterns at the undesired angles, which in part lead to higher levels of interference introduced to the received subcarriers in those angles. As the objective function explicitly depends on the radiation patterns at different time steps, and since more variations in the radiation patterns are desirable for satisfying the above requirements, the higher number of unit cells $N$ would result in higher levels of security and reliability. It is important to note that one could add the phase delay values, i.e., $\beta_0p$ and $\beta_1p$, to the list of optimization variables in the above problem as well. In particular, as $\beta_0p$ can be easily controlled by modifying the input voltage to the varactors of each unit cell, one could also optimize the above problem over the possible values of $\beta_0p$. Generally, higher feasible values for phase delays are desirable as it makes a wider range of beam scanning possible for the LWA's radiation patterns. We also comment on the role of the number of time steps ($L$) in the above optimization problem. We recall that the reliability constraint calls for the radiation patterns at different time steps to have the same value. Therefore, $L=2$ can be deemed as the most suitable choice for meeting the reliability constraint as it is more challenging to satisfy this constraint for larger $L$'s. However, one would have higher degrees of freedom for controlling the level of interference in the undesired angles by increasing $L$. In fact, one could compromise on the reliability level for the desired angle by increasing $L$ in order to ensure a higher security level at the undesired angles.

\begin{figure}
\centering
\hspace*{-1cm}      
\includegraphics[width=1.05\textwidth, height=16cm]{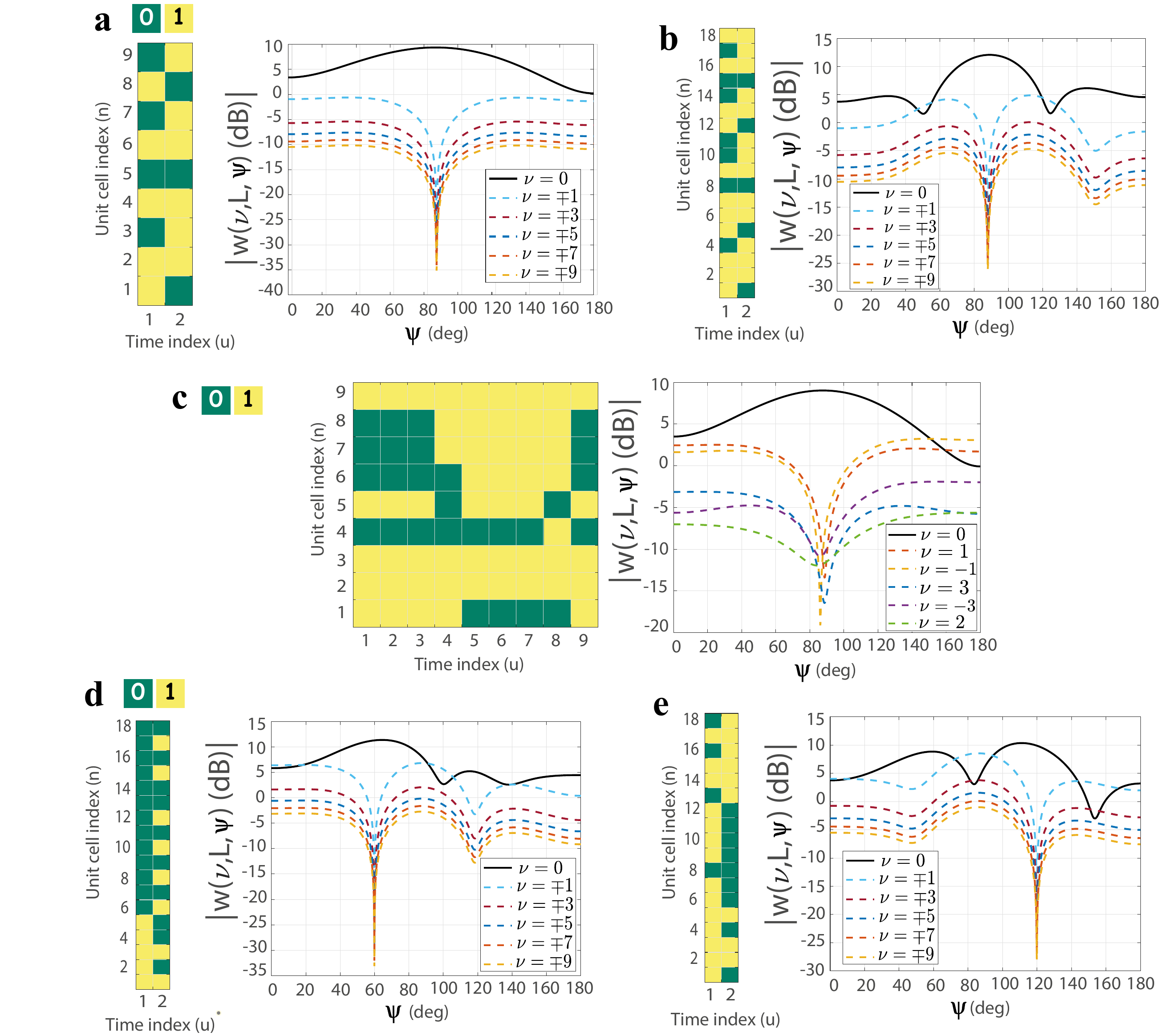}
\caption{Illustration of the dominant harmonic patterns generated by the proposed $1D$ space-time digitally-coded MTM-LWA along with the corresponding digital codes obtained from solving (\ref{eq:optimizationproblem_freespace}). The green and yellow squares represent $0$ and $1$, respectively. \textbf{a} Number of unit cells, number of time steps, and the desired angle are set to $N=9$, $L=2$, and $\psi_0=88^\circ$, respectively. \textbf{b} Effect of higher number of unit cells $N=18$ in the generated harmonics. \textbf{c} Effect of increasing the number of time step to $L=9$. \textbf{d}, \textbf{e} Achieving DM for different desired angles, $\psi_0=60^\circ$ and $\psi_0=120^\circ$, by solely modifying the digital codes.}
\label{fig:All-1D}
\end{figure}

The stated problem in Equation \eqref{eq:optimizationproblem_freespace} is known to be MINLP optimization problem where the objective function is non-linear and the optimization variables are a mix of integer and real-valued quantities. We have also verified that the Hessian of the above objective function could have negative eigenvalues in general, which makes the problem non-convex. This kind of problem is NP-hard and no efficient global optimal solver is available in the literature \cite{B&B2}. A well-known mathematical optimization technique for finding local minima to the above problem is the branch and bound algorithm \cite{BranchandBound}, which is the state-of-the-art solver for MINLPs, and we utilize it to find the digital coding sequences as elaborated in the Methods section.

Figure \ref{fig:All-1D} illustrates pattern of the generated harmonics from a $1$D space-time digitally-coded MTM-LWA whose coding sequence is obtained by solving the above MINLP for different parameters $L$, $N$ and $\psi_0$, where $\beta_0p=-18^\circ$, $\beta_1p=15^\circ$, $p=1.5$ cm, and $f_0=1.95$ GHz. Specifically, Figure \ref{fig:All-1D}.a represents the harmonic patterns for $N=9$ and $L=2$. This figure indicates that the coding sequences result in a very small value for all the harmonics at the desired angle $\psi_0=88^\circ$ while the level of harmonics increases as one gets further away from $\psi_0$. This indeed means that the received constellation will not be interfered with in the desired angle while it would be perturbed in the other angles. Figure \ref{fig:All-1D}.b shows that increasing the number of unit cells not only improves antenna directivity with a narrower main beam but also effectively increases the level of higher-order harmonics in undesired directions, leading to more interference for the signal received by Eve. Furthermore, as will be demonstrated in Figures \ref{fig:Bers-All}.a and \ref{fig:Bers-All}.b, increasing the number of unit cells results in a more rapid increase in BER as one moves further from the desired angle compared to the case with a lower $N$, resulting in higher levels of security against Eve. On the other hand, increasing the number of time steps $L$, as illustrated in Figure \ref{fig:All-1D}.c, increases the level of harmonics in undesired directions at the expense of increasing them in the desired direction as well, representing a trade-off between reliability and security. Another important functionality of our proposed architecture is illustrated in Figures \ref{fig:All-1D}.d and \ref{fig:All-1D}.e  where it is shown that one is able to achieve DM for different given desired angles by using the proper coding sequences obtained from solving the above MINLP. This would indeed reduce the complexity of the overall architecture compared to the existing time-modulated phased-arrays systems achieving DM \cite{OFDM_DM,OurACM}, as it circumvents the need for the phase shifters.

\section{2D space-time digitally-coded MTM-LWA array}

 \begin{figure}
\centering
\includegraphics[width=1\textwidth,height=9.4cm]{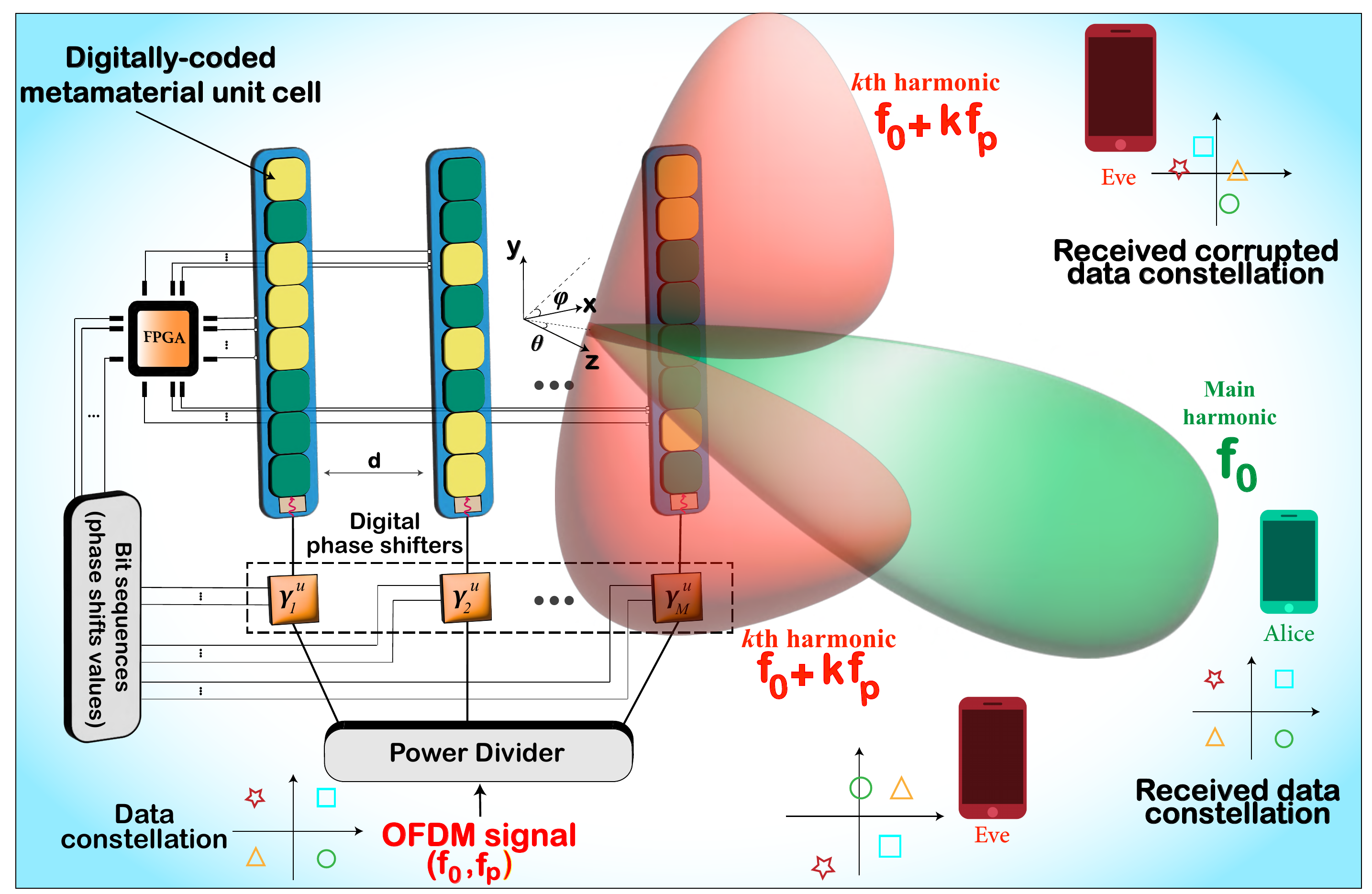}
\caption{Demonstration of the proposed space-time digitally-coded LWA in $2$D fashion. Using this architecture, one is able to steer the main beam direction corresponding to the fundamental frequency towards the desired $2$D direction, at the same time suppressing the received power in the other harmonic frequencies in that direction. This leads to preservation of the data constellation for each OFDM subcarrier in the direction of the legitimate receiver while distorting it in the other directions.}
\label{fig:ProposedSchemes2}
\end{figure}

The above space-time digitally-coded MTM-LWA can also be extended to the 2D setting by incorporating a parallel feeding architecture at the transmitter (Tx) consisting of $M$-element linear antenna array as shown in Figure \ref{fig:ProposedSchemes2}. In this case, each antenna branch is equipped with a digital phase shifter followed by a space-time digitally coded MTM-LWA which enables $2$D beam-scanning functionality. The radiation pattern for this architecture is given by
\begin{align}
\label{eq:RadiationPattern2D}
R(\theta,\phi,t)=\sum_{m=1}^M\sum_{n=1}^N S(t) e^{-\alpha(n-1)p}e^{j(n-1)k_0p\sin{\theta}\sin{\phi}}e^{j(m-1)k_0d\sin{\theta}\cos{\phi}}U_{mn}(t),
\end{align}
where $\theta$ and $\phi$ denote the elevation and azimuth receiving angles, respectively. Also. the phase-delay function for this architecture is given by
\begin{align}
U_{mn}(t)=\sum_{u=1}^L \Gamma_{nm}(\mathbf{q}_u)\Lambda_m^uH^u(t),
\end{align}
for $\Lambda_m^u= e^{-j\gamma_m^u}$, $\gamma_m^u=(m-1)k_0d\sin{\theta_0^u}\cos{\phi_0^u}$, and $\Gamma_{nm}(\mathbf{q}_u)= \prod_{k=1}^ne^{j\kappa_{\mathbf{q}_{u}(k,m)}}$ where $t\in\mathcal{T}_p$ and $\theta_0$ and $\phi_0$ denote the desired elevation and azimuth angles, respectively. We note that $\Lambda_m^u$ is the phase shifter value of the $m$th branch during $\mathcal{T}^u$. As digital phase shifters have finite precision in practice, the amount of phase shift should be set to the closest quantized value to $\Lambda_m^u$. Also, similar to LWAs' states, phase shifters' values change periodically with a period of $\frac{1}{f_p}$ which is governed by binary sequences generated by an FPGA. Focusing on the case in which two states are available for each unit cell in each branch, the digital coding sequence corresponding to the $k$th unit cell in the $m$th branch is given by $\kappa_{\mathbf{q}_{u}(k,m)}=\begin{cases}
 \beta_0p,~{\mathbf{q}_{u}(k,m)}=0,\\
 \beta_1p,~{\mathbf{q}_{u}(k,m)}=1,
 \end{cases}$ during $\mathcal{T}^u$.

  The radiation pattern in this case can be obtained in a similar fashion to the above $1$D digitally-coded MTM-LWA as $R(\theta,\phi,t)=S(t)\sum_{\nu=-\infty}^\infty w(\nu,L,t,\theta,\phi)$ where
\begin{align}
\label{eq:w_function_2D}
    w(\nu,L,t,\theta,\phi)=e^{j2\pi\nu f_pt}\frac{1}{L}\text{sinc}\big(\frac{\nu\pi}{L}\big)e^{\frac{j\pi\nu}{L}}\sum_{u=1}^L\Xi(\theta,\phi,\mathbf{q}_u)e^{-j\frac{2\nu\pi u}{L}},
\end{align}
and
\begin{align}
\label{eq:Xi_2D}
    \Xi(\theta,\phi,\mathbf{q}_u)=\sum_{m=1}^M\sum_{n=1}^N\Gamma_{nm}(\mathbf{q}_u)\Lambda_m^u e^{-\alpha(n-1)p}e^{j(n-1)k_0p\sin{\theta}\sin{\phi}}e^{j(m-1)k_0d\sin{\theta}\cos{\phi}}.
\end{align}

Similar to the 1D case, we then mathematically formulate the problem of finding digital coding sequences satisfying the three DM constraints mentioned in the previous section for a given $\theta_0$ and $\phi_0$ as (see Section S$3$ of the Supporting Information) 
\begin{align}
\label{eq:optimizationproblem_freespace_2D}
    \argmin_{\substack{\mathbf{q}_u,d_u^1,d_u^2\\u=1,\dots,L}}~ \sum_{u=2}^L \big|&\Xi(\theta_0,\phi_0,\mathbf{q}_u)-\Xi(\theta_0,\phi_0,\mathbf{q}_1)\big|-\sum_{u=1}^L\big|\Xi(\theta_0+d_u^1,\phi_0+d_u^2,\mathbf{q}_u)\big|\nonumber\\&
    \text{s.t.}~~~\mathbf{q}_u\in\{0,1\}^{M\times N},~ L_u\leq d_u^1,~d_u^2\leq U_u,
\end{align}
where $\theta_0$ and $\phi_0$ denote the desired angles corresponding to the direction of legitimate receiver. Also, $d_u^1$ and $d_u^2$ are the optimization variables determining the amount of deviation of the radiation pattern from the desired angles at time step $\mathcal{T}_u$. We note that the coding sequences, $\mathbf{q}_u$, in this case is a matrix whose rows correspond to the digital codes for the LWA in each branch. 

\begin{figure}
\hspace*{-1.2cm} 
\centering
\includegraphics[width=1.2\textwidth,height=13cm]{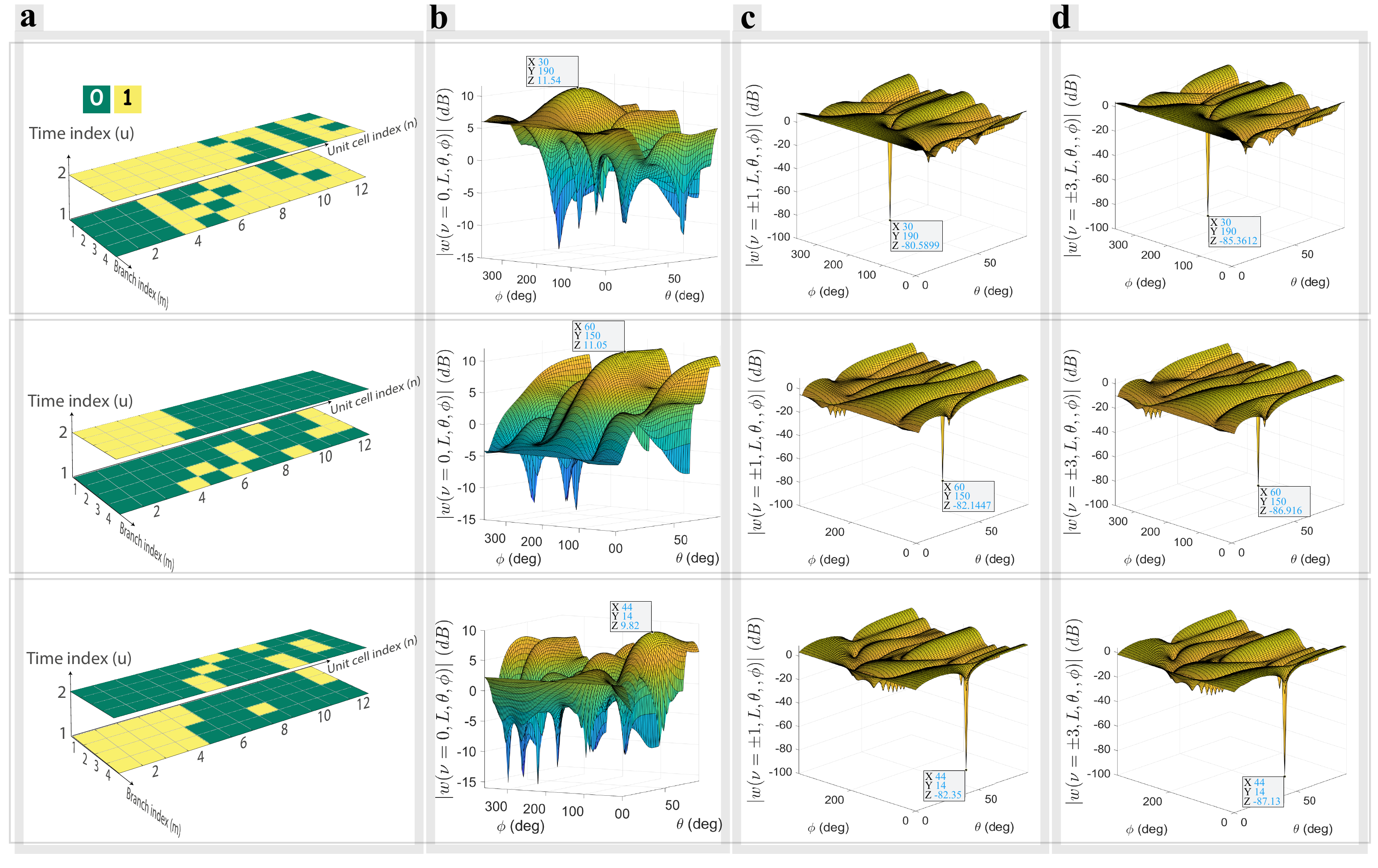}
\caption{Resulting harmonic patterns from the proposed $2$D MTM-LWA when $N=12$, $M=4$, and $L=2$ for three different sets of desired angles, i.e., ($\theta_0=30^\circ$, $\phi_0=190^\circ$), ($\theta_0=60^\circ$, $\phi_0=150^\circ$), and ($\theta_0=44^\circ$, $\phi_0=14^\circ$), each corresponding to a row. \textbf{a} The coding sequences obtained as a result of solving the MINLP. \textbf{b} Patterns of the fundamental frequency ($\nu=0$), i.e., $w(\nu=0,L,\theta,\phi)$. \textbf{c} Pattern of the first positive and negative harmonics. \textbf{d} Pattern of the third positive and negative harmonics.}
\label{fig:All-2D}
\end{figure}

Figure \ref{fig:All-2D} extends our results to the $2$D case pertaining to the architecture in Figure \ref{fig:ProposedSchemes2} for the case where $\beta_0p=-18^\circ$, $\beta_1p=15^\circ$, $p=1.5$ cm, $f_0=1.95$ GHz, and $d$ is set to be half of the wavelength. For the figures in the first row, the coding sequences are obtained to achieve DM for the desired angles $\theta_0=30^\circ$ and $\phi_0=190^\circ$ while we have $\theta_0=60^\circ$, $\phi_0=150^\circ$ and $\theta_0=44^\circ$, $\phi_0=14^\circ$ for the second and third row, respectively. Figure \ref{fig:All-2D}.a demonstrates the coding sequences for each case obtained via solving the MINLP in Equation \eqref{eq:optimizationproblem_freespace} for the $2$D case. Figure \ref{fig:All-2D}.b displays the harmonic pattern of the fundamental frequency ($\nu=0$), which shows that the space-time digitally-coded MTM-LWA transmits the maximum power {toward the desired direction} with the designed coding sequences. Figures \ref{fig:All-2D}.c and \ref{fig:All-2D}.d, on the other hand, illustrate the pattern of the first and third positive and negative harmonics in the $2$D space as the dominant harmonics contributing the most to the spectral interference. We note that we have not shown even-numbered harmonics here as they have been suppressed, i.e., their values are close to zero, and their contribution is negligible as a result. It is shown that the level of harmonics is extremely low for the desired angles while it rapidly increases for every other angle. This is the result of the designed coding sequences for MTM-LWAs and phase shift values in each branch for an $8$-bit digital phase shifter which leads to DM in the $2$D space.

\section{Wireless channel and BER analysis}
Until now we have shown the applicability of our proposed digitally coded MTM-LWAs for enabling DM in free space where there is only line of sight communication between the transmitter and the receiver. We now consider communications through a wireless channel where there could be more than one transmission path between the communication parties. This so-called multi-path effect is caused by refraction and reflection of waves from water bodies and terrestrial objects. The knowledge of channel gains between the antenna elements at the transmitter and those at a receiver, known as channel state information (CSI), is commonly available at two levels of perfect instantaneous CSI and partial CSI \cite{Sina_Alireza_survey2019,Sina_MIMO}, where the former amounts to the actual realization of the channel gain at a given time, while the latter implies only the statistical distribution of the channel. In practice, CSI knowledge can readily be acquired in OFDM systems where certain subcarriers are allotted for sending/receiving known signals termed `pilot'. Having access to the received version of the pilot and the actual transmitted pilot, one can utilize methods like least squares to estimate the channel coefficients \cite{Channel_Estimation}. Given the CSI of the Bob's and Eve's channels, Alice can perform beamforming in order to ensure data can be reliably decoded by Bob while preventing Eve from doing so. Such beamforming can be efficiently performed in the proposed programmable MTM-LWAs through the use of digital codes. The mathematical expression for the received signal in this case, which is a function of channel gains between the antenna elements at the transmitter and those at the receiver, can be obtained in a similar fashion to Equation \eqref{eq:RadiationPattern2D} based on the $\Xi$ function in Equation \eqref{eq:Xi_2D}, which here we denote by $\Xi(\mathbf{H},\mathbf{q}_u)$ to emphasize its dependence on the channel gains $\mathbf{H}$ (see Section S$4$ of the Supporting Information).

As already alluded to, the digital codes in this case are found based on the criteria of reliability for Bob and security against Eve. As shown in Section S$4$ of the Supporting Information, the optimization problem can be formulated as follows

\begin{align}
\label{eq:optimization_problem_channel}
    \argmin_{\substack{\mathbf{q}_u\\u=1,\dots,L}}~ \sum_{u=2}^L \big|&\Xi(\mathbf{H}^{AB},\mathbf{q}_u)-\Xi(\mathbf{H}^{AB},\mathbf{q}_1)\big|-\sum_{u=2}^L\big|\Xi(\mathbf{H}^{EB},\mathbf{q}_u)-(\mathbf{H}^{EB},\mathbf{q}_1)\big|\nonumber\\&
    ~~~~~~~~~~~\text{s.t.}~~~~\mathbf{q}_u\in\{0,1\}^{1\times N},
\end{align}
where $\mathbf{H}^{AB}$ and $\mathbf{H}^{EB}$ denote the instantaneous CSI corresponding to Alice-Bob and Eve-Bob channels, respectively. The first summation imposes the requirement that the radiation patterns' values computed based on the CSI of Alice-Bob channel must be similar for all the time steps. Meanwhile, the difference between those values given the CSI of Eve-Bob channel is made as large as possible through the second summation term. In simple words, the first term removes the interference terms for Bob by canceling the effect of the channel $\mathbf{H}^{AB}$. The second summation, on the other hand, maximizes the contribution of the interference terms to the received signal through the channel $\mathbf{H}^{EB}$. For the cases where only the partial CSI of Eve-Bob channel is available at Alice, a similar optimization problem can be formulated which includes averaging over different realizations of the channel (Section S$4$ of the Supporting Information). Similar to the free space scenario, these optimization problems fall into the category of mixed integer non-linear programs which can efficiently be solved via branch and bound algorithm. We note that the channel gains are time-varying in general which necessitates solving Equation \eqref{eq:optimization_problem_channel} whenever channel gains change. We further discuss this matter in the next section.

\begin{figure}
\centering
\includegraphics[width=\textwidth, height=19cm]{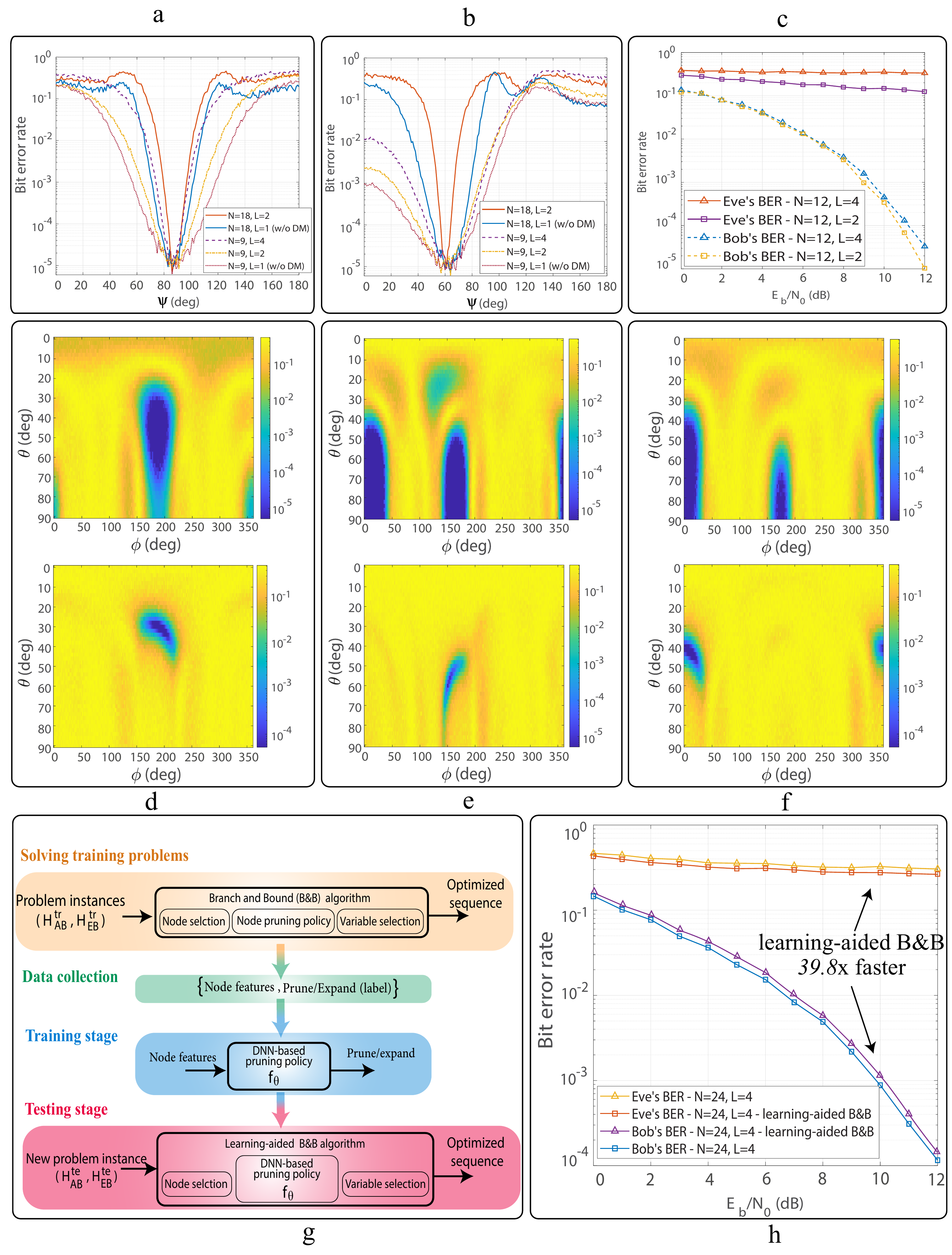}
\caption{BER results corresponding to the proposed space-time digitally-coded MTM-LWA architecture. For the $2$D cases, the number of unit cells and number of branches are set to $N=12$ and $M=4$, respectively. \textbf{a} $1$D case where the desired angle is set to $\psi_0=88^\circ$. \textbf{b} $1$D case for $\psi_0=60^\circ$. \textbf{c} Bob's and Eve's BER in a wireless channel scenario as a function of the receiving SNR. \textbf{d} BER's heatmap for the $2$D case where $\theta_0=30^\circ$, and $\phi_0=190^\circ$. Upper: without DM, lower: time-modulated with $L=2$.  \textbf{e} BER's heatmap for the $2$D case where $\theta_0=60^\circ$, and $\phi_0=150^\circ$. Upper: without DM, lower: time-modulated with $L=2$. \textbf{f} BER's heatmap for the $2$D case where $\theta_0=44^\circ$, and $\phi_0=14^\circ$. Upper: without DM, lower: time-modulated with $L=2$. \textbf{g} A flow-chart of learning-aided B\&B algorithm for speeding up the optimization process. \textbf{h} Performance of the learning-aided B\&B algorithm compared to the original B\&B algorithm in terms of BER rate seen by Bob and Eve.}
\label{fig:Bers-All}
\end{figure}

So far, we have mainly evaluated the performance of the proposed space-time digitally-coded MTM-LWA in terms of the harmonics' patterns. However, in real-world communication systems the reliability of a communication link is widely measured in terms of BER. For a given direction of the legitimate receiver, we evaluate the BER performance across the entire $1$D (or $2$D) space by sweeping the angles and computing the BER for each direction. By examining the corresponding angles in the presented results, we can extract the observed BERs for both Bob and Eve. A BER value close to zero indicates a reliable link, while a value around $0.5$ suggests that the receiver is performing no better than random guessing of the information bits. We present the BER performance for our proposed transmitter's architectures in Figure \ref{fig:Bers-All}, and compare it with the architectures without DM in a variety of scenarios. Figures \ref{fig:Bers-All}.a and \ref{fig:Bers-All}.b demonstrate the results of the $1$D case for $\psi_0=88^\circ$ and $\psi_0=60^\circ$, respectively, for different parameters. The curves labeled as ``w/o DM" correspond to the scenario in which the antennas are only space-modulated, i.e., without time modulation, resulting in no generation of harmonic interference at the receiver. In this case, the coding sequences are obtained to ensure that the angle of the main beam coincides with that of the fundamental tone in the corresponding DM cases.
  
One can observe that larger number of unit cells and time steps would improve the overall system performance in terms of providing reliability for Bob and security against Eves, which indeed translates to a narrower BER curve. The BER results for the $2$D case are illustrated in Figures \ref{fig:Bers-All}.d, \ref{fig:Bers-All}.e, and \ref{fig:Bers-All}.f for different desired receiving angles. In these figures, the upper heatmap corresponds to the architecture without DM while the lower one demonstrates the BER's heatmap for the time-modulated case. The effectiveness of our proposed architecture to achieve DM's functionalities is highlighted in all these three figures from the BER perspective. Finally, Figure \ref{fig:Bers-All}.c pertains to the BER performance of Bob and Eve over a wide range of SNR per bit ($E_b/N_0$) where a wireless channel is considered between the communication parties for two sets of parameters, i.e., $N=12,~L=2$ and $N=12,~L=4$ . In particular, perfect CSI of Alice-Bob channel is assumed to be known at Alice while only partial CSI of Eve-Bob channel is given to her. It is shown that when SNR is high a reliable communication link with low BER is provided to Bob with the designed coding sequences, while preventing Eve from correctly decoding the information bits for a large span of SNR. Here, one can also observe that a higher number of time steps would result in higher BERs for Eve at the expense of a less reliable link for Bob.

\section{Enhancing solver's speed via deep learning}
In a dynamic wireless setting where either the communication parties themselves or their surrounding objects are not spatially fixed, the channel gains are expected to vary over time. The duration in which the channel coefficients remain constant is commonly known as the coherence time of the channel. The implication of this is that the optimization problem in Equation \eqref{eq:optimization_problem_channel} needs to be solved for each coherence time associated with the Alice-Bob and Alice-Eve channels. Therefore, time efficiency is an important requirement for the solver to be able to find the digital codes for the current channel gains in a timely manner. The B\&B algorithm, as the state-of-the-art technique for solving MINLPs, has exponential computational complexity \cite{B&B2} as the number of integer optimization variables grows. This makes the B\&B algorithm impractical for our purpose of finding temporal digital codes in real time as the time complexity of the solver could be prohibitive given the coherence times of wireless channels under consideration.

The B\&B algorithm finds the solution by iteratively searching a binary tree based on the MINLP. At each node, the integrality constraints are dropped and a relaxed version of the optimization problem is solved whose objective value serves as a lower bound to the original problem. A global upper bound is only available when a feasible solution, which satisfies all the constraints including the integrality ones, is found at the node. The algorithm pops a node from a stack of nodes, evaluates the aforesaid two bounds, and then decides whether to prune a node or proceed to explore its children. This process continues until the node stack is empty. The three main components of the branch and bound algorithm during this searching process are node selection, variable selection, and node pruning policy \cite{B&B4}. The high time complexity of the branch and bound algorithm is attributed to the computational complexity of these three main components. In particular, node pruning policy is of utmost significance as it controls the size of the node stack. The original pruning policy of the algorithm takes the most conservative approach aimed at finding the optimal solution and guaranteeing its optimality by investigating all the feasible solutions. By designing an alternative pruning policy, one can significantly reduce the computational complexity of the algorithm by pruning the unpromising portions of the search tree more aggressively at the cost of compromising on a certificate for optimality \cite{B&B4,B&B2}.

We leverage supervised and transfer learning methods to devise a new node pruning policy for the B\&B algorithm, which we refer to as learning-aided B\&B. During the offline training stage, multiple instances of the problem are solved for different CSIs. Although the parameters of the optimization problem are different, they maintain the same binary search tree structure which can be exploited by a supervised learning algorithm. Specifically, while solving the training problems with the B\&B algorithm, training data representing important features pertaining to the pruning process are collected at each node. These features are of two different types: structural features and CSI features. The former correspond to the structure of the underlying binary search tree consisting of features related to a node at a local level or the whole tree at a global level. The CSI features, as the name implies, are related to the specific channel gain information at a node level. We train a deep neural network (DNN) classifier, as the node pruning policy, on the training data which, given the feature vector as the input, decides whether a particular node ought to be pruned or not. As detailed in the Methods, the training of this DNN is done over different stages. At each stage, the DNN trained on the previous stage is used as the pruning policy to collect further training data. Then, we utilize fine-tuning method \cite{meroune,HyPhyLearn,MLSP}, which is a transfer learning technique, to update the weights of the DNN given the new training data. Figure \ref{fig:Bers-All}.g illustrates a flow-chart of the learning-aided B\&B algorithm where a DNN-based pruning policy is trained and utilized in the original B\&B algorithm. Figure \ref{fig:Bers-All}.h demonstrates the performance of the learning-aided B\&B algorithm when $N=24$ and $L=4$ in terms of BER at Bob and Eve. In the training phase the optimization problems are solved for data collected for $40$ different CSIs from each Alice-Bob and Alice-Eve channels. 
It should be noted that the size of the training data used to train the DNN-based pruning policy shown in Figure \ref{fig:Bers-All}.g depends on the number of nodes visited by the B\&B algorithm to reach a solution. Each entry in the training dataset includes the features of a node and its corresponding label, as illustrated in Figure \ref{fig:Bers-All}.g. As a result, the size of the training dataset is directly proportional to the number of nodes visited by the B\&B algorithm. Depending on the number of unknowns in the problem, a large number of nodes may be visited before a solution is reached, resulting in a dataset that captures the pruning policy's behavior in the B\&B algorithm.
Moreover, it is worth mentioning that our learning-aided B\&B approach achieved a speedup of $39.8$ times compared to the original algorithm for the case of $N = 24$ and $L = 4$, which demonstrates the effectiveness of our approach in accelerating the search process. The speed is measured in terms of the number of visited nodes before reaching a final solution. Using a commercial laptop with a $4$-core Intel
processor and $8$ GB of RAM, the time consumption for the learning-aided B\&B algorithm for the case of $N=12$ and $L=2$ is around $1.2$ s while for the original B\&B method takes around $4$ minutes to find a solution. The cost paid for such speed enhancement is compromising on the achieved reliability and security levels. However, we point out that the BER performance degradation, as shown in Figure \ref{fig:Bers-All}, is not significant in terms of both system reliability/security, while the resulting speed improvement is substantial.

\section{Experimental verification}
\begin{figure}
\centering
\hspace*{-3cm} 
\includegraphics[width=1.1\textwidth, height=16cm]{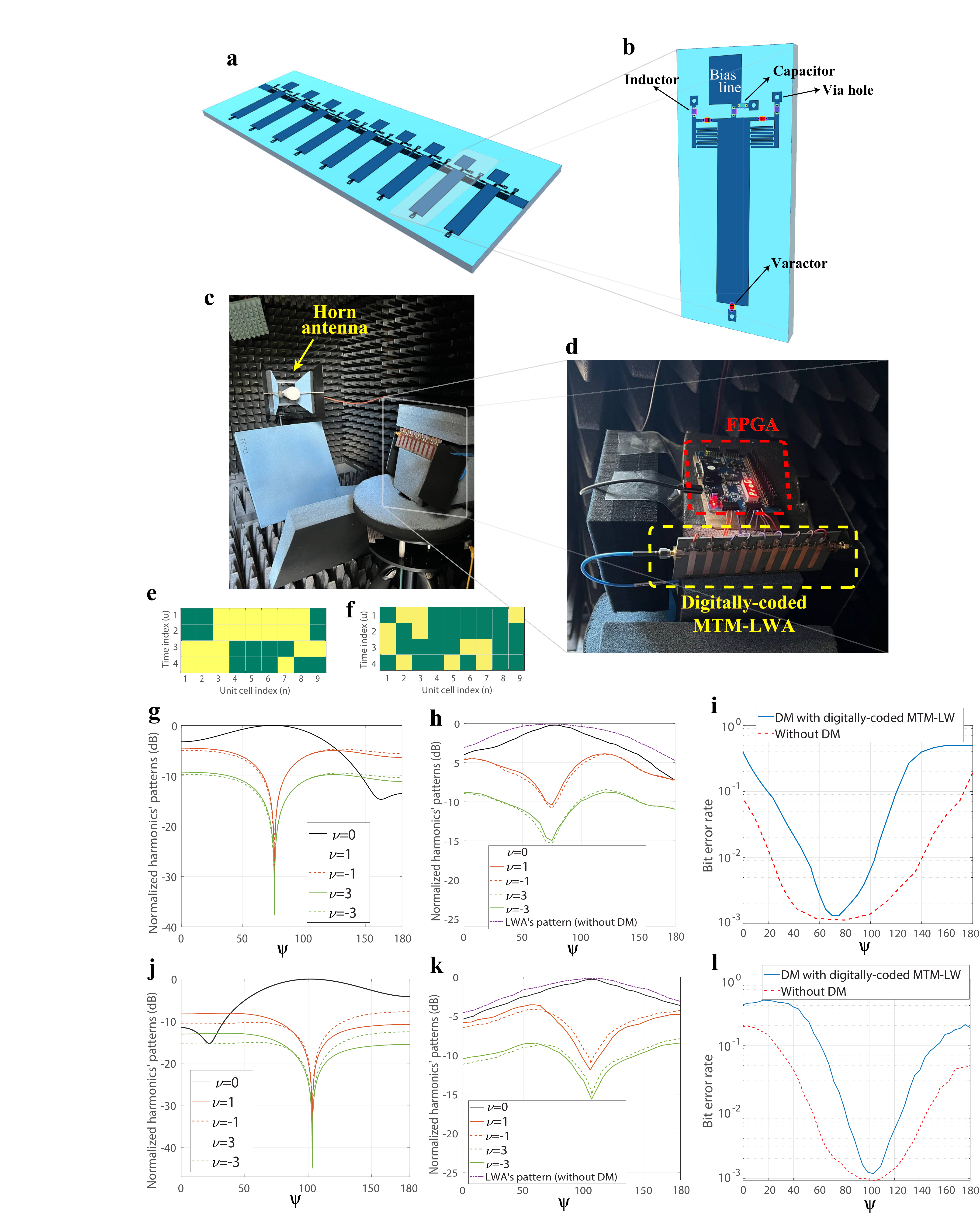}
\caption{Prototype design and experimental verification. \textbf{a} Schematic of the realized space-time digitally-coded MTM-LWA with $9$ unit cells. \textbf{b} Schematic of the tunable unit cell with the incorporated varactors. \textbf{c} Measurement setup in an anechoic chamber. \textbf{d} Fabricated prototype. \textbf{e} The optimized coding sequences for the desired angle $\psi_0=75^\circ$. \textbf{f} The optimized coding sequences for the desired angle $\psi_0=105^\circ$. \textbf{g} Numerically computed normalized harmonic patterns for $\psi_0=75^\circ$. \textbf{h} Measured normalized harmonic patterns for $\psi_0=75^\circ$. \textbf{i} BER measurement results corresponding to the with and without DM transmissions for the case of $\psi_0=75^\circ$. \textbf{j} Numerically computed normalized harmonic patterns for $\psi_0=105^\circ$. \textbf{k} Measured normalized harmonic patterns for $\psi_0=105^\circ$. \textbf{l} BER measurement results corresponding to the with and without DM transmissions for the case of $\psi_0=105^\circ$.}
\label{fig:Experimental_All}
\vspace{-0.25in}
\end{figure}

In order to validate the DM functionality of the proposed architecture, we realize a space-time digitally-coded MTM-LWA with $9$ unit cells pertaining to the $1$D transmitter architecture in Figure \ref{fig:ProposedSchemes}.a. A schematic of the designed CRLH LWA is illustrated in Figure \ref{fig:Experimental_All}.a. Figure \ref{fig:Experimental_All}.b shows a blow-up version of the unit cell consisting of interdigital capacitors, shunt stub, and three varactors sharing a common bias voltage to make the unit cell tunable similar to the design procedure reported in \cite{1381686}. The length of each unit cell ($p$) is $1.52$ cm. We apply two levels of bias voltages to varactors via FPGA, each making the underlying unit cell work in either state ``$0$" ($\beta_0p<0$) or state ``$1$" ($\beta_1p>0$) in the frequency around $2$ GHz located in the fast wave region of the unit cell (see Section S$6$ of the Supporting Information for the dispersion diagram of the unit cells).
As shown in Figure \ref{fig:Experimental_All}.d, a prototype of the LWA with $9$ tunable CRLH cells was fabricated on a RO5880 board with dielectric constant of $2.2$ and thickness of $1.57$ mm. In accordance with the measurement setup illustrated in Figure \ref{fig:Experimental_All}.c, we have carried out the experiments in a standard microwave anechoic chamber. The prototype is mounted on a turntable, which can automatically rotate by
$180^\circ$ in the horizontal plane. A horn antenna is used to
receive the propagated signal (in fundamental and harmonics frequencies) by the prototype. The dynamic bias voltages, equivalent to coding sequences, for the varactors of each unit cell are provided through an FPGA board (BASYS 3) with a clock speed of $100$MHz, in which a code is preloaded to generate nine digital control voltages according to the
time-coding sequences in Figure \ref{fig:Experimental_All}.e where $L=4$. The switching period $T_p$ of the time-coding sequences is $630$ ns which corresponds to the modulation frequency of $1.58$ MHz. To measure the fundamental and harmonic patterns, an input signal with the frequency $1.95$ GHz is injected to the sample and a spectrum analyzer connected to the horn antenna located in the far-field region of the prototype records harmonic levels in different angles. We have solved the corresponding optimization problem for achieving DM for two different desired angles of $\psi_0=75^\circ$ and $\psi_0=105^\circ$ on two different sides of the broadside angle. According to the dispersion diagram of the metamaterial unit cell, which is provided in Section S$6$ of the Supporting Information, one could verify that $\beta_0p=-18^\circ$ and $\beta_1p=15^\circ$, which leads to the coding sequence provided in Figures \ref{fig:Experimental_All}.e and \ref{fig:Experimental_All}.f for the $\psi_0=75^\circ$ and $\psi_0=105^\circ$ case, respectively. These sequences are generated and fed to the prototype through the FPGA. For the case of $\psi_0=75^\circ$, Figure \ref{fig:Experimental_All}.g depicts the numerically computed normalized pattern of the main and $4$ most dominant harmonics, while Figure \ref{fig:Experimental_All}.h presents their corresponding pairs measured by the spectrum analyzer connected to the horn antenna. Similarly, Figures \ref{fig:Experimental_All}.j and \ref{fig:Experimental_All}.k illustrate the simulated and measured harmonic patterns for the $\psi_0=105^\circ$ case, respectively. For both cases, one can verify the DM functionalities for the desired angle by comparing the theoretical and measurement results. In fact, measured results show that the main beam direction of the fundamental frequency is the same as desired angle $\psi_0$ in which the received power in the higher-order harmonic frequencies is minimum and at least $10$ dB less than that in the fundamental frequency.

It should be noted that the coding sequences used in our study are theoretically expected to suppress even-numbered harmonics, which is consistent with the measurement results showing that the level of these harmonics falls more than $30$ dB below the dominant ones. Therefore, in Figures \ref{fig:Experimental_All}.h and \ref{fig:Experimental_All}.k, we have chosen not to display the even-numbered harmonics to emphasize the dominant ones. This similarity between the theoretical and measured results in harmonic patterns supports the concept of space-time LWA. However, we note that the null levels in the simulation results are much lower than those observed in the measurements. This can be attributed to non-ideal time-varying characteristics of the components, such as varactors, and also to a realistic wireless environment. The simulation results are based on the array factor approach, as explained in Section $2$, which does not account for these factors.

For the BER measurements, the transmitted OFDM signal with a carrier frequency of $1.95$ GHz, subcarrier width of $15$ kHz, and a total number of $64$ subcarriers is generated via a commercially available software-defined radio (SDR) module. QPSK modulation is employed for mapping the information bits to the constellation points. Another SDR is also used to process the signal received by the horn antenna. Specifically, GNU Radio interface is used to implement the IEEE $802.11$ standard for transmission and reception of OFDM packets. We have measured the BER through sending/receiving OFDM packets over the air (see Section S$5$ of the Supporting Information for more details). These results are illustrated in Figures \ref{fig:Experimental_All}.i and \ref{fig:Experimental_All}.l for the $\psi_0=75^\circ$ and $\psi_0=105^\circ$ cases, respectively. For one to appreciate this BER performance, it is necessary to compare it against a transmitter without DM. To this end, we have fixed the states of the cells in our MTM-LWA to $[0~ 0~ 1~ 1~ 1~ 1~ 0~ 0~ 0]$ and $[0~ 0~ 0~ 1~ 0~ 0~ 0~ 0~ 1]$ whose radiation patterns are illustrated in Figures \ref{fig:Experimental_All}.h and \ref{fig:Experimental_All}.k, respectively. As shown, the main beam directions for these two cases are the same as that of their corresponding fundamental frequency pattern in the time-modulated 
case, while the beam is more directive for the latter case due to DM. It can be seen that by using the DM scheme one can substantially improve the security of the system in physical layer as the BER in the undesired angles is much higher than that of the not time-modulated case. Also, note that under the proposed scheme the DM can be achieved for two different desired angles by the mere change of the coding sequences. Compared to the similar $1$D DM schemes, the proposed space-time digitally-coded MTM-LWA substantially reduces the complexity as it circumvents the need for any phase shifters or antenna feed network design.

\section{Discussion}
In conclusion, we have proposed a DM scheme based on the idea of space-time digitally-coded-modulated MTM-LWA, in which propagation constant of each unit cell is periodically controlled by a set of temporal coding sequences. We have presented theoretical results concerning the radiation pattern in $1$D and $2$D cases for digitally-coded MTM-LWA. We have shown that by carefully designing the coding sequences, one can achieve the functionalities of the DM, i.e., providing reliability for the intended receiver while making the system secure against the eavesdroppers. To this end, an optimization problem is formulated which takes into account both spatial and spectral requirements pertaining to the DM. Furthermore, we have shown that the proposed transmitter architecture can be used to provide DM in a wireless channel setting where perfect or partial CSI is available at the transmitter. Extensive provided simulation results confirm the applicability of the proposed approach for achieving DM as a way of enhancing physical layer security. Compared to other existing DM transmitter, our newly-proposed architecture reduces the hardware complexity as it does not rely on phase shifters or an antenna feeding structure. As a proof of concept, we have further fabricated a digitally-coded CRLH LWA prototype in which an FPGA controller provides the predetermined coding sequences to the unit cells. This experimental setup have helped us to verify our presented theory in practice through both harmonic patterns and BER measurements.

\section{Methods}
\subsection{Branch and bound algorithm as the solver for MINLP}
The B\&B algorithm solves a (mixed) integer non-linear program by iteratively searching a binary tree in which each node $n$ is associated with a subproblem of the original problem. For each subproblem, constraints of the integer variables are modified in a way that some integer variables are determined while others are undetermined and relaxed into continuous variables within $[0,1]$ for the binary case. Solving the corresponding subproblem at node $n$ results in a local lower bound, $L_n$, to the original solution, as the feasible region of the subproblem is larger than that of the original MINLP problem. The searching process of the B\&B algorithm contains four iterative steps. 
\begin{enumerate}[1)]
    \item Node selection: selecting a node from the unvisited node list of the tree. 
	\item Variable selection: selecting an optimization variable from the list of variables for the current node.
	\item Evaluation: solving the corresponding nonlinear subproblem of the node to obtain its local upper bound.
	\item Pruning policy: This determines whether the current node is worth expanding or not. Specifically, using the local lower bound, $L_n$, and the global upper bound, $U$, which is the optimal value of the objective function by far, the algorithm decides whether to prune a node or explore its children. The searching process comes to an end by iteratively repeating the above three steps until the node list is empty.
\end{enumerate}

For simplicity, we adopt the depth-first-search as the node selection rule and always choose the first undetermined element in the coding sequence vector for variable selection process. The original prune policy of the B\&B algorithm includes three cases: 
\begin{enumerate}[1)]
	\item The sub-problem is infeasible: If the relaxed nonlinear sub-problem in node $n$ is infeasible, the related MINLP problem is also infeasible and then the node $n$ is pruned. 
	 \item A feasible solution is found: If the current solution in node $n$ is an integral vector, the result is also a feasible solution of the related MINLP problem and then the node $n$ is fathomed. 
	 \item The local lower bound is greater than the current global upper: In this case, the node $n$ would not lead to a better solution and is fathomed.
\end{enumerate}

\subsection{Enhancing speed of the B\&B algorithm with deep learning}
Here, we elaborate on the speed enhancement procedure utilized in this paper for the B\&B algorithm. We first note that the B\&B algorithm aims for two primary goals, i.e., finding a solution and guaranteeing its optimality. The latter goal is achieved by searching all feasible solutions and comparing the found solution with them. Most of the time is consumed in this latter goal and a more aggressive pruning policy can significantly reduce the computational complexity. In fact, by pruning more nodes, the underlying search space is being limited, resulting in less time being consumed. The price that is being paid for this increase in speed is that the solutions cannot be claimed to be optimal. Although the resulting solution would be suboptimal, the performance gap might be negligible compared to the achieved gain in speed. To this end and inspired by the literature of B\&B algorithm, we aim for designing an alternative pruning policy that would prune all the non-optimal nodes. The speed enhancement of the B\&B algorithm consists of the following steps. We begin by solving a number of training examples via the original B\&B algorithm and collect training data in the form of node features and the corresponding labels, which shows whether the B\&B algorithm has pruned a particular node or not. The node features correspond to the structure of binary tree utilized by the B\&B algorithm and also the specific information about the CSIs. As the structural (problem independent) features, we have used the depth of node $n$, the plunge depth of node $n$, its local lower bound $L_n$, value of branching variable, current global upper bound, and the number of solutions found so far. Also, CSI information from Alice-Bob and Alice-Eve channels pertaining to the current branching variable is also stored among the node features. Then, a DNN with three hidden layers, each with $200$ number of neurons, is trained on the collected dataset, which given a node feature vector, predicts a binary label $0$ (not prune) or $1$ (prune). We train this neural network in multiple rounds each of which corresponding to one training example. Once the DNN is initially trained based on the dataset associated with the first training example, in each of the following rounds we utilize fine-tuning \cite{meroune,HyPhyLearn} as a transfer learning technique to refine the DNN’s weights. In this way, the DNN learns to mimic the punning policy of the B\&B algorithm over different rounds for multiple channel realizations.

\subsection{BER measurements with SDRs}
The BER measurement procedure is detailed as follows. Within a laptop with Ubuntu operating system, the GNU radio interface is used to implement the IEEE $802.11$ standard based on OFDM packets. In total $500000$ bits are transmitted in packets of length $25$. The transmitted bits are saved in a file. Then, the generated OFDM signals are inputted to our proposed prototype, i.e., digitally-coded space-time leaky wave antenna, via a SDR and subsequently gets transmitted over the air. At the reception side, a horn antenna is used to received the time-modulated OFDM signals. Another SDR is used to process the received signal via the GNU radio interface and save the decoded bits in a new file. By comparing the two saved files the BER is calculated for the over the air communication. A schematic of the BER measurement is provided in Section S$5$ of the Supporting Information.

\section*{Data availability}
The data that support the findings of this study are available from the corresponding author upon reasonable request.

\section*{Acknowledgements}
This work was supported by the National Science Foundation (NSF) under Grant ECCS-2229384 and ECCS-2028823. Any opinions, findings, and conclusions or recommendations expressed in this material are those of the author(s) and do not necessarily reflect the views of the National Science Foundation.

\section*{Author contributions}

Initial conceptualization: C.-T.M.W.; methodology (algorithm $\&$ prototyping), experiments and data collection: A.N., S.V.; technical investigation and feedback: W.U.B., N.B.M.; supervision: C.-T.M.W., W.U.B., N.B.M; Writing —original draft: A.N., S.V.; writing—review $\&$ editing: C.-T.M.W, W.U.B., N.B.M. All the authors discussed results and commented on the manuscript.

\section*{Supporting Information}
Supporting Information is available from the Wiley Online Library or from the author.

\section*{Competing interests}
The authors declare no competing interests.
\section*{Materials $\&$ Correspondence}

Correspondence and requests for materials should be addressed to C.-T.M.W.

\newpage
\setcounter{figure}{0}
\renewcommand{\figurename}{Figure}
\renewcommand{\thefigure}{S\arabic{figure}}
Supplementary Information for
\\
\\
{\huge Programming Wireless Security through Learning-Aided
Spatiotemporal Digital Coding Metamaterial Antenna}
\\
\\
 Alireza Nooraiepour, Shaghayegh Vosoughitabar, Chung-Tse Michael Wu, Waheed U. Bajwa, and Narayan B. Mandayam  

\vspace{8em}
\hspace{-15pt}\textbf{Supplementary Note 1: Derivation of Fourier-series coefficients}
\(\mathbf{c}_{\mathbf{\nu n}}\)
\\

In the following, we obtain the Fourier series coefficients of the periodic
function \(U_{n}(t)\) with fundamental frequency of
\(f_{p} = \frac{1}{T_{p}}\).

\[c_{\nu n} = \frac{1}{T_{p}}\int_{0}^{T_{p}}{U_{n}(t)e^{- j2\pi\nu f_{p}t}dt} = \frac{1}{T_{p}}\int_{0}^{T_{p}}{\sum_{u = 1}^{L}{\Gamma_{n}\left( \mathbf{q}_{u} \right)H^{u}(t)}e^{- j2\pi\nu f_{p}t}dt}\]

\[= \frac{1}{T_{p}}\sum_{u = 1}^{L}{\Gamma_{n}\left( \mathbf{q}_{u} \right)}\int_{\frac{(u - 1)T_{p}}{L}}^{\frac{uT_{p}}{L}}e^{- j2\pi\nu f_{p}t}dt \]

\[= \sum_{u = 1}^{L}{\Gamma_{n}\left( \mathbf{q}_{u} \right)\frac{1}{T_{p}}}{\frac{- 1}{j2\pi\nu f_{p}}e}^{- j2\pi\nu f_{p}t}\Bigg|_{\frac{(u - 1)T_{p}}{L}}^{\frac{uT_{p}}{L}}\]

\[= \sum_{u = 1}^{L}{\Gamma_{n}\left( \mathbf{q}_{u} \right)\frac{- 1}{j2\pi\nu}\left( e^{- \frac{j2\pi\nu u}{L}} - e^{- \frac{j2\pi\nu(u - 1)}{L}} \right)}\]

\[= \sum_{u = 1}^{L}{\Gamma_{n}\left( \mathbf{q}_{u} \right)\frac{- 1}{j2\pi\nu}e^{- \frac{j2\pi\nu u}{L}}\left( 1 - e^{- \frac{j2\pi\nu}{L}} \right)}\]

\[= \sum_{u = 1}^{L}{\Gamma_{n}\left( \mathbf{q}_{u} \right)\frac{- 1}{j2\pi\nu}e^{- \frac{j\pi\nu(2u - 1)}{L}}\left( e^{- \frac{j\pi\nu}{L}} - e^{\frac{j\pi\nu}{L}} \right)}\]

\[= \sum_{u = 1}^{L}{\Gamma_{n}\left( \mathbf{q}_{u} \right)\frac{- 1}{j2\pi\nu}e^{- \frac{j\pi\nu(2u - 1)}{L}}\left( - 2j\sin\frac{\pi\nu}{L} \right)}\]

\[= \sum_{u = 1}^{L}{\Gamma_{n}\left( \mathbf{q}_{u} \right)\frac{\text{sinc}(\pi\nu/L)}{L}e^{- \frac{j\pi\nu(2u - 1)}{L}}}\]

\hspace{-15pt}\textbf{Supplementary Note 2: Derivation of the received signal in a
given direction}
\\

\begin{itemize}
\item
  \textbf{1D case:}
\end{itemize}

If the constraints \(\left\{ \begin{matrix}
w\left( \nu \neq 0,L,t,\psi_{0} \right) = 0 \\
w\left( \nu = 0,L,t,\psi_{0} \right) \neq 0 \\
\end{matrix} \right.\ \) are satisfied, the received signal
\(R\left( {\psi_{0}}_{\ },t \right)\ \)at the desired angle
\(\psi_{0}\) would reduce to

\[R\left( {\psi_{0}}_{\ },t \right) = S(t)w\left( \nu = 0,L,t,\psi_{0} \right) = \frac{1}{KL}\sum_{u = 1}^{L}{\Xi^{u}\left( \theta_{0},\mathbf{q}_{u} \right)}\sum_{k = 1}^{K}{s_{k}e^{j2\pi\left( f_{c} + (k - 1)f_{p} \right)t}},\]

which is a weighted version of the transmitted OFDM signal.

For the undesired angles where
\(w\left( \nu,L,t,{\psi \neq \psi_{0}}_{\ } \right) \neq 0,\ \forall\nu\),
the received signal at the subcarrier \(f_{0} + xf_{p}\) is given by

\[R\left( {\psi_{\ }},t \right)\Big{|}_{f_{0} + xf_{p}} = \sum_{k = 1}^{K}{s_{k}e^{j2\pi\left( f_{c} + (k - 1)f_{p} \right)t}}w\left( \nu = x - k,L,t,\psi_{\ } \right)\]

which is a function of the complex data from all the subcarriers and
as well as the interference terms
\(w\left( \nu = x - k,L,t,\psi_{\ } \right)\).

\begin{itemize}
\item
  \textbf{2D case:}
\end{itemize}

Similar to the 1D case, given that the constraints
\(\left\{ \begin{matrix}
w\left( \nu \neq 0,L,t,\theta_{0},\phi_{0} \right) = 0 \\
w\left( \nu = 0,L,t,\theta_{0},\phi_{0} \right) \neq 0 \\
\end{matrix} \right.\ \) are satisfied for the desired angles
\(\theta_{0}\), \(\phi_{0}\), the received signal
\(R\left( {\theta_{0}}_{\ },{\phi_{0}}_{\ },t \right)\ \)equals to

\[R\left( {\theta_{0}}_{\ },\phi_{0},t \right) = \sum_{u = 1}^{L}{\Xi^{u}\left( \theta_{0},\phi_{0},\mathbf{q}_{u} \right)}\sum_{k = 1}^{K}{s_{k}e^{j2\pi\left( f_{c} + (k - 1)f_{p} \right)t}},\]
while the received signal at the tone \(f_{0} + xf_{p}\) for the other
directions would be corrupted by the interference terms as follows:

\[R(\theta,\phi,t)\Big{|}_{f_{0} + xf_{p}} = \sum_{k = 1}^{K}{s_{k}e^{j2\pi\left( f_{c} + (k - 1)f_{p} \right)t}}w\left( \nu = x - k,L,t,\theta_{\ },\phi \right).\]

\vspace{3em}
\hspace{-15pt}\textbf{Supplementary Note 3: Problem formulation for DM in free space}
\\

First, we note that the received signal in the 2D case can be stated as
follows\footnote{The problem formulation follows similar steps for the
  1D case.}:
\[R(\theta,\phi,t) = S(t)\sum_{\nu = - \infty}^{\infty\ }{e^{j2\pi\nu f_{p}t}\frac{1}{L}\text{sinc}\left( \frac{\nu\pi}{L} \right)e^{\frac{j\pi\nu}{L}}\sum_{u = 1}^{L}{\Xi\left( \theta,\phi,\mathbf{q}_{u} \right)e^{\frac{j2\nu\pi u}{L}}}}\ \]
where \(\Xi\left( \theta,\phi,\mathbf{q}_{u} \right)\) is given in
equation (10) of the paper. The first optimization constraint pertaining
to the desired direction can be stated by \(\left\{ \begin{matrix}
\frac{1}{L}\text{sinc}\left( \frac{\nu\pi}{L} \right)e^{\frac{j\pi\nu}{L}}\sum_{u = 1}^{L}{\Xi\left( \theta_{0},\phi_{0},\mathbf{q}_{u} \right)e^{\frac{j2\nu\pi u}{L}}} = 0 \\
\frac{1}{L}\sum_{u = 1}^{L}{\Xi\left( \theta_{0},\phi_{0},\mathbf{q}_{u} \right)} \neq 0 \\
\end{matrix} \right.\ \). Here, the requirement on the second line is
always satisfied while the first one imposes that
\(\sum_{u = 1}^{L}{\Xi\left( \theta_{0},\phi_{0},\mathbf{q}_{u} \right)e^{\frac{j2\nu\pi u}{L}}} = 0\).
This condition is equivalent to
\(\Xi\left( \theta_{0},\phi_{0},\mathbf{q}_{u_{i}} \right) = \Xi\left( \theta_{0},\phi_{0},\mathbf{q}_{u_{j}} \right),\ \forall u_{i},u_{j}\)
since it always holds that the sum of the \(L\)th roots of unity equals to
zero, i.e., \(\sum_{u = 1}^{L}e^{\frac{j2\nu\pi u}{L}} = 0\). The second
optimization constraint boils down to
\(\sum_{u = 1}^{L}{\Xi\left( \theta_{\ },\phi_{\ },\mathbf{q}_{u} \right)e^{\frac{j2\nu\pi u}{L}}} \neq 0,\ \forall\nu,\theta,\ \phi\)
which is satisfied by imposing that
\(\Xi\left( \theta_{\ },\phi_{\ },\mathbf{q}_{u_{i}} \right) \neq \Xi\left( \theta_{\ },\phi_{\ },\mathbf{q}_{u_{j}} \right),\ \forall u_{i},u_{j},\theta,\ \phi\).
Finally, the third optimization constraint corresponding to the MTM-LWA
beam pattern can be stated as \({\theta_{0}}_{\ },{\phi_{0}}_{\ } = \text{argmax}_{\theta,\ \phi}\ \left| \Xi\left( \theta_{\ },\phi_{\ },\mathbf{q}_{u_{\ }} \right) \right|,\ \forall u\).
As \(\Xi\left( \theta_{\ },\phi_{\ },\mathbf{q}_{u_{\ }} \right)\)
corresponds to a LWA radiation pattern, this latter constraint results
in time-invariant (not time-modulated) coding sequences, i.e.,
\(\mathbf{q}_{u_{i}} = \mathbf{q}_{u_{j}},\ \forall u_{i},\ u_{j}\),
which do not generate any harmonics. Therefore, aiming at satisfying
the second constraint and achieving directional modulation, we relax the
third constraint as
\({\theta_{0}}_{\ },{\phi_{0}}_{\ } = \text{argmax}_{\theta,\ \phi}\ \left| \Xi\left( {\theta + d_{u}^{1}}_{\ },{\phi + d_{u}^{2}}_{\ },\mathbf{q}_{u_{\ }} \right) \right|,\ \forall u\)
by introducing new variables \(d_{u}^{1}\) and \(d_{u}^{2}\). This
formulates the optimization problem as follows:

\[\underset{\mathbf{q}_{u},d_{u}^{1},d_{u}^{2}}{\text{arg}\min}{\sum_{u = 2}^{L}\left| \Xi\left( {\theta_{0}}_{\ },{\phi_{0}}_{\ },\mathbf{q}_{u_{\ }} \right) - \Xi\left( {\theta_{0}}_{\ },{\phi_{0}}_{\ },\mathbf{q}_{1} \right) \right|} - \sum_{u = 1}^{L}\left| \Xi\left( {\theta_{0} + d_{u}^{1}}_{\ },{\phi_{0} + d_{u}^{2}}_{\ },\mathbf{q}_{u_{\ }} \right) \right|\]
\hspace{12em}s.t.
\(\mathbf{q}_{u_{\ }} \in \left\{ 0,1 \right\}^{M \times N},\ L_{u} \leq d_{u}^{1},d_{u}^{2} \leq U_{u},\)

where \(L_{u}\) and \(U_{u}\) correspond to the lower and upper limits
for the introduced variables, respectively.

\begin{figure}[!h]
\centering   
\includegraphics[width=0.8\textwidth]{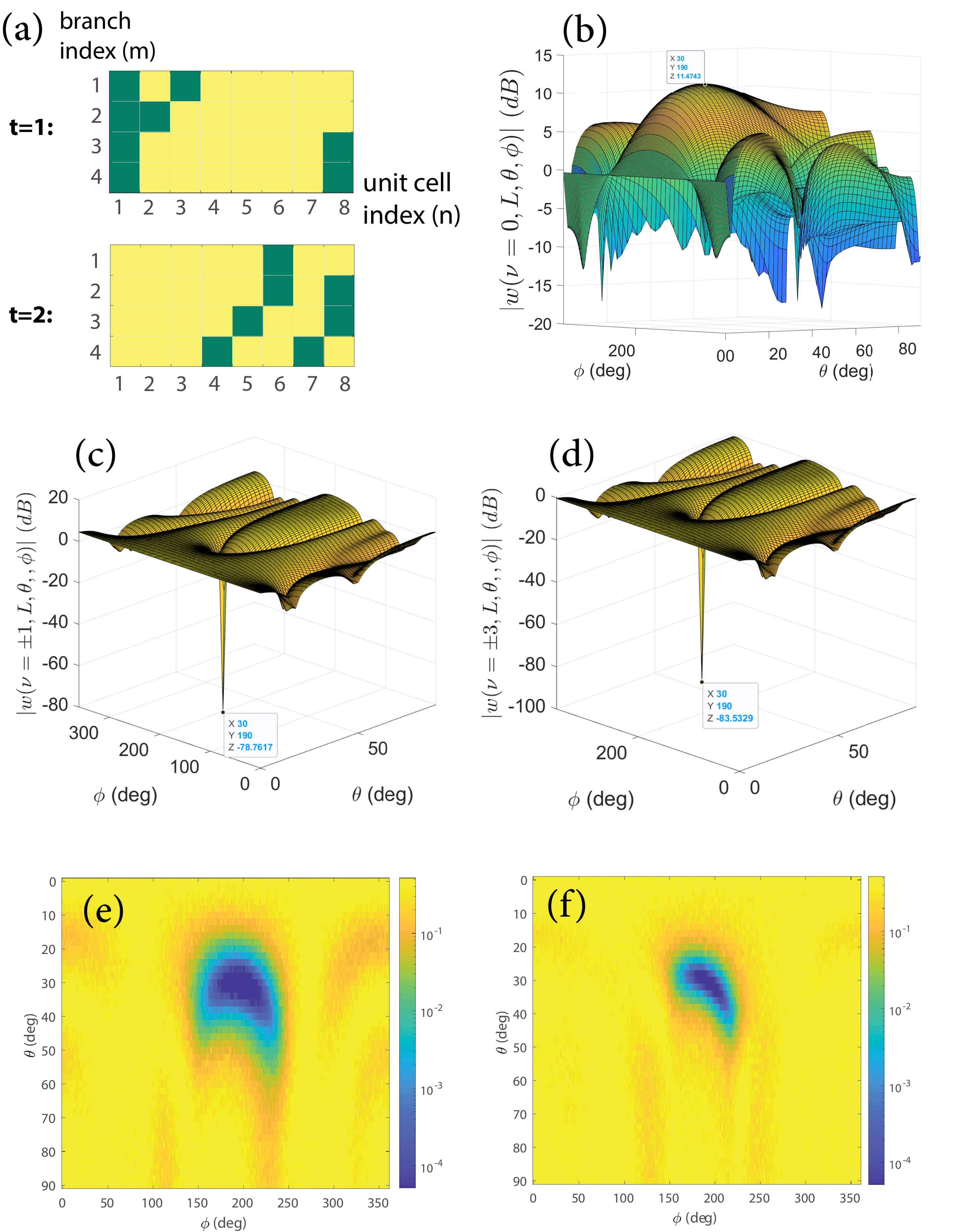}
\caption{Resulting harmonic patterns from the proposed $2$D MTM-LWA when $N=8$, $M=4$, and $L=2$ for the desired angles $\theta_0=30^\circ$, $\phi_0=190^\circ$. \textbf{a} The coding sequences obtained as a result of solving the MINLP. \textbf{b} Harmonic pattern of the fundamental frequency ($\nu=0$), i.e., $w(\nu=0,L,\theta,\phi)$. \textbf{c} Pattern of the first positive and negative harmonics. \textbf{d} Pattern of the third positive and negative harmonics.
\textbf{e} Resulting BER in the $(\theta,\phi)$ space from the  $N=8$, $M=4$, and $L=2$ scheme. \textbf{f} Resulting BER in the $(\theta,\phi)$ space from the $N=12$, $M=4$, and $L=2$ scheme which is presented in the manuscript.}
\label{Fig:N8L2_190and30}. 
\end{figure}

In the remaining, we have conducted further investigation for the 2D case. Specifically, we have obtained coding sequences for the case where the legitimate receiver is located in the direction $\theta_0=30^\circ$ and $\phi_0=190^\circ$, assuming there are $M=4$ branches, each containing $N=8$ unit cells and number of time steps $L=2$. Fig. \ref{Fig:N8L2_190and30}.a - \ref{Fig:N8L2_190and30}.d illustrate the coding sequences, main harmonic, and the first four dominant higher-order harmonics. By comparing these results to the $N=12$ case, which is presented in the paper, we have seen that higher-order harmonic levels are around $4$ dB higher in average. Therefore, similar to the $1$D case, increasing the number of unit cells leads to raising the harmonic levels for the undesired angles. This indeed causes more distortion for the received signal by Eve at the undesired angles or equivalently higher BER for Eve. Specifically, this can be verified by comparing the corresponding BER results from the $N=8$ and $N=12$ cases presented in Fig. \ref{Fig:N8L2_190and30}.e, \ref{Fig:N8L2_190and30}.f, respectively, where the latter leads to a narrower low BER region (higher security) due to using a greater number of unit cells. We further note that increasing the number of time steps $L$ would have similar effect to 1D, which is raising the harmonics levels for the undesired angles, i.e., higher security against Eve, but at the same time raising the null level, which leads to lower reliability for the legitimate receiver.

\hspace{-15pt}\textbf{Supplementary Note 4: Problem formulation for DM in the case of
wireless channel}
\\

The received signal from a digitally-coded LWA in the presence of a
wireless channel is obtained by

\[R(t) = \sum_{n = 1}^{N}{\sum_{k = 1}^{K}{s_{k}e^{j2\pi\left( f_{c} + (k - 1)f_{p} \right)t}}}e^{- \alpha(n - 1)p}h_{nk}^{AB}U_{n}(t)\]

where the number of receive antenna is set to 1 without loss of
generality. The periodic phase-delay function is defined as
\(U_{n}(t) = \sum_{u = 1}^{L}{\Gamma_{n}\left( \mathbf{q}_{u} \right)H^{u}(t)}\), the same way as equation (2) in the paper. In particular, we note that
the above signal is normalized for the path loss and
\(h_{nk}^{AB}\ \)represents the normalized complex channel gain between
Alice and Bob. Similar to the free space scenario, by expanding
\(U_{n}(t)\) in terms of its Fourier series we get

\[R^{AB}(t) = \sum_{k = 1}^{K}{s_{k}e^{j2\pi\left( f_{c} + (k - 1)f_{p} \right)t}}\sum_{m = - \infty}^{\infty}{e^{j2\pi mf_{p}t}\frac{1}{L}\text{sinc}\left( \frac{m\pi}{L} \right)e^{\frac{j\pi m}{L}}}\]

\[\times \left\lbrack \sum_{u = 1}^{L}{\left( \sum_{n = 1}^{N}{\Gamma_{n}\left( \mathbf{q}_{u} \right)}e^{- \alpha(n - 1)p}h_{nk}^{AB} \right)e^{- \frac{j2m\pi u}{L}}\ }_{\ } \right\rbrack.\]

For the case of narrow-band communications, the channel gains for an
antenna element over the whole frequency band can be taken to be identical,
i.e., \(h_{ni} \approx h_{nj},\ \forall i,j\). For this case, we define

\[\Xi(\mathbf{H}^{AB},\mathbf{q}_{u}) = \sum_{n = 1}^{N}{\Gamma_{n}\left( \mathbf{q}_{u} \right)}e^{- \alpha(n - 1)p}h_{n}^{AB},\]

where $\mathbf{H}^{AB}$ denotes the instantaneous CSI vector consisting of $h_n^{AB}$'s. For the Alice-eavesdropper (AE) link, the received signal $R^{AE}(t)$ can be similarly obtained as above, given the corresponding instantaneous CSI $\mathbf{H}^{AE}$:

\[\Xi(\mathbf{H}^{AE},\mathbf{q}_{u}) = \sum_{n = 1}^{N}{\Gamma_{n}\left( \mathbf{q}_{u} \right)}e^{- \alpha(n - 1)p}h_{n}^{AE}\]

In order to provide DM towards the legitimate receiver, i.e.,
reliability for Bob and security for Eve, we again invoke the fact that
the sum of the \(L\)th roots of unity equals to zero. Subsequently, we
can formulate the problem as follows

\[\underset{\mathbf{q}_{u}}{\text{arg}\min}{\sum_{u = 2}^{L}\left| \Xi\left( \mathbf{H}^{AB},\mathbf{q}_{u_{\ }} \right) - \Xi\left( \mathbf{H}^{AB},\mathbf{q}_{1} \right) \right|} - \sum_{u = 2}^{L}\left| \Xi\left( \mathbf{H}^{AE},\mathbf{q}_{u_{\ }} \right) - \Xi\left( \mathbf{H}^{AE},\mathbf{q}_{1} \right) \right|\]
\hspace{12em}s.t. \(\mathbf{q}_{u_{\ }} \in \left\{ 0,1 \right\}^{1 \times N}.\)
\\

For the scenario where only partial CSI of the Alice-Eve link is
available, we note that the AE channel can be modeled as a doubly
correlated fading MIMO channel \cite{Sina_MIMO}, namely,
\(H^{AE} = \psi_{r}^{1/2\ }\widehat{H_{e}}\psi_{t}^{1/2\ },\) where
\(\psi_{r}^{\ \ }\) and \(\psi_{t}^{\ }\) are the receive and transmit
correlation matrices, respectively. \(\widehat{H_{e}}\) is a complex
matrix with independent and identically distributed zero mean unit
variance circularly symmetric complex Gaussian entries. Partial CSI of
AE channel corresponds to only the statistics of the channel, i.e.,
knowledge of correlation matrices. For this scenario, the goal of DM
would be to cancel spectral interferences for Alice while ensuring
eavesdroppers with channel statistics
\(H^{AE} = \psi_{r}^{1/2\ }\widehat{H_{e}}\psi_{t}^{1/2\ }\) are
corrupted by these interferences. To this end, we generate \(N_{t}\)
number of realizations for \(\widehat{H_{e}}\) which corresponds to
\(N_{t}\) instances for the AE channel denoted by
\(H^{AE,i},\ i = 1,\ldots,N_{t}\). Then the objective function for the
DM optimization consists of an averaging term corresponding to all the
\(N_{t}\) realizations, as indicated in the following:
\[\underset{\mathbf{q}_{u}}{\text{arg}\min}{\sum_{u = 2}^{L}\left| \Xi\left( \mathbf{H}^{AB},\mathbf{q}_{u_{\ }} \right) - \Xi\left( \mathbf{H}^{AB},\mathbf{q}_{1} \right) \right|} - \frac{1}{N_{t}}\sum_{i = 1}^{N_{t}}{\sum_{u = 2}^{L}\left| \Xi\left( \mathbf{H}^{AE,i},\mathbf{q}_{u_{\ }} \right) - \Xi\left( \mathbf{H}^{AE,i},\mathbf{q}_{1} \right) \right|}\]
\hspace{12em}s.t. \(\mathbf{q}_{u_{\ }} \in \left\{ 0,1 \right\}^{1 \times N}\).
\vspace{3em}

In the following table we have provided the BER degradation values corresponding to the learning-aided algorithm that is trained based on data corresponding to $40$ CSIs from each of the Alice-Bob and Alice-Eve channels. The following table summarizes these results for two different set of ($N,L$) parameters and various SNR values. For the case of ($N=12,L=2$), the degradation values are smaller as the search space for the branch and bound algorithm is limited compared to the ($N=24,L=4$) case, where it is easier for the learning-aided algorithm to mimic the node pruning policy of the original branch and bound. Also, focusing on the ($N=24,L=4$) case and assuming a working SNR of $12$ dB for both Bob and Eve, one can confirm from the Figure $5$.h in the original manuscript that even the degraded BER values corresponding to the learning-aided algorithm still results in low BER of around $10^{-4}$ for Bob and high BER of around $0.3$ for Eve.

\begin{table}[!h] 
\scriptsize
 \centering
\label{Table: PerformanceDegredation}
\begin{tabular}{|l|c|c|c|c|c|c|c|}
\hline
\diagbox[width=8em]{Case study}{SNR (dB)} & $0$&$2$   & $4$ &$6$ & $8$& $10$& $12$ \\ \hline
\begin{tabular}[c]{@{}l@{}}Bob's error\\ ($N=24$, $L=4$)\end{tabular}
& $\num{2e-2}$ & $\num{1.2e-2}$  & $\num{1.2e-2}$& $\num{2e-3}$ &  $\num{1.2e-3}$ & $\num{1.8e-4}$ & $\num{4.4e-5}$\\ \hline
\begin{tabular}[c]{@{}l@{}}Eve's error\\ ($N=24$, $L=4$)\end{tabular} & $\num{2e-2}$ &  $\num{2.2e-2}$ & $\num{2.5e-2}$ & $\num{2.8e-2}$&  $\num{2.8e-2}$ & $\num{3.2e-2}$ & $\num{3.5e-2}$\\ \hline
\begin{tabular}[c]{@{}l@{}}Bob's error\\ ($N=12$, $L=2$)\end{tabular}  & $\num{1.4e-3}$ &  $\num{2e-3}$ & $\num{2e-3}$& $\num{4.2e-3}$&  $\num{2.2e-4}$ & $\num{2e-5}$ & $\num{2.5e-6}$ \\ \hline
\begin{tabular}[c]{@{}l@{}}Eve's error\\ ($N=12$, $L=2$)\end{tabular}  & $\num{1.6e-2}$ &  $\num{1.6e-2}$ & $\num{2e-2}$ & $\num{2e-3}$&  $\num{2.2e-3}$ & $\num{2.6e-3}$ & $\num{2.8e-2}$\\ \hline
\end{tabular}
    \end{table}
                
\begin{figure}[!h]
\centering   
\includegraphics[width=1\textwidth]{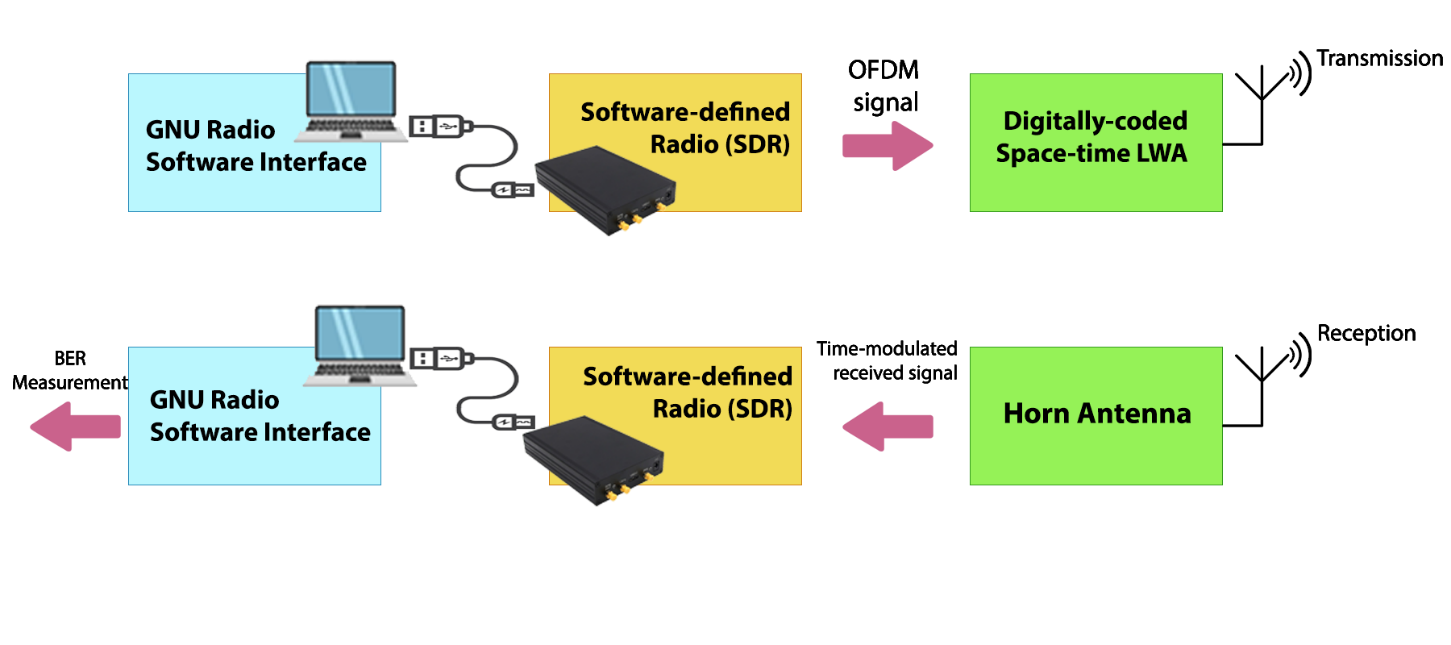}
\caption{BER measurement procedure conducted using the SDRs.}
\label{Fig:SDR}. 
\end{figure}

\hspace{-15pt}\textbf{Supplementary Note 5: BER measurements with software-defined radios (SDRs)}
\\
Fig. \ref{Fig:SDR} illustrates a schematic of the BER measurement procedure that is used to obtain BER results by sending/receiving bits over the air. As discussed in the methods OFDM signals are generated from the GNU radio interface which carry information bits over the air with the help of SDR from the transmitter to the receiver. At the reception side, a horn antenna is used to received the time-modulated OFDM signals and decode the information bits. 

\begin{figure}[h]
\centering   
\includegraphics[width=0.7\textwidth]{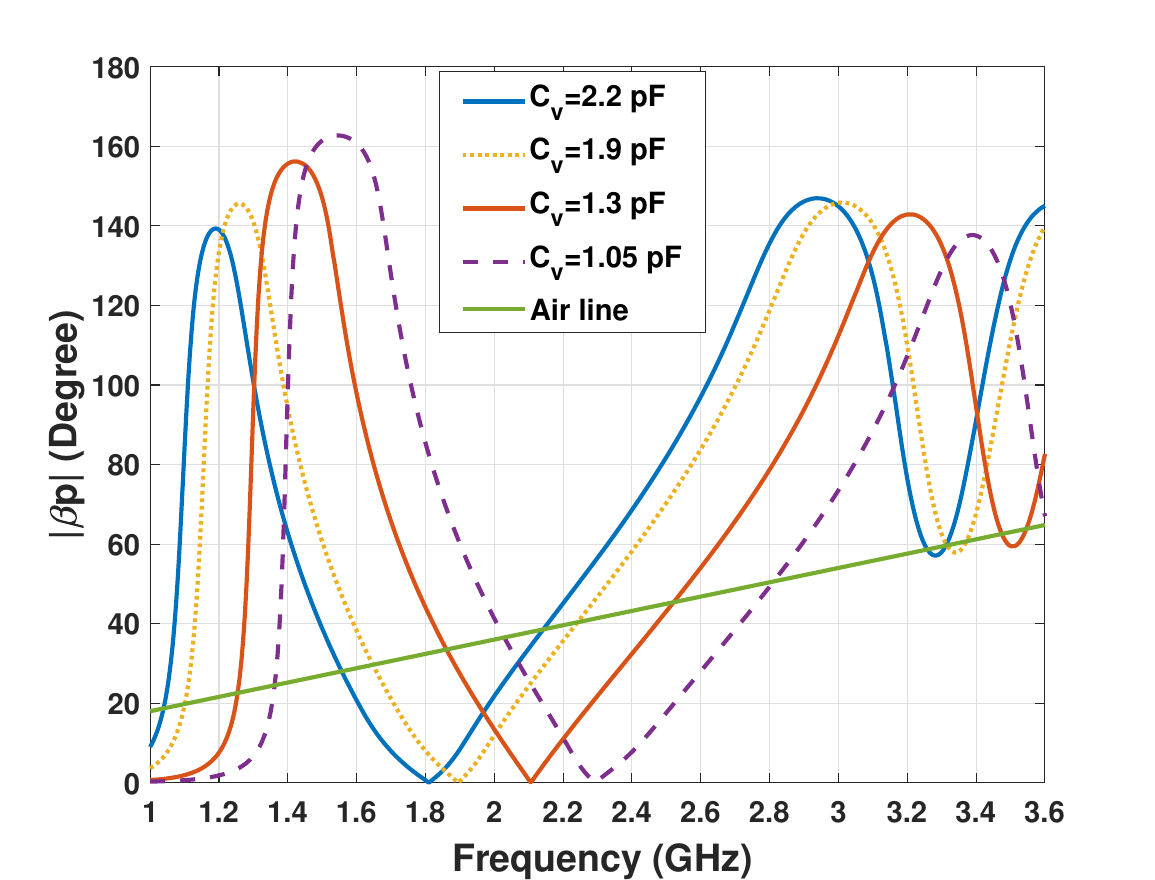}
\caption{Dispersion diagrams of the tunable CRLH unit cell for different varactor values based on the simulation in HFSS.}
\label{Fig:Dispersion}
\end{figure}

\hspace{-15pt}\textbf{Supplementary Note 6: Dispersion diagrams of the tunable CRLH unit cell}
\\
In Fig. \ref{Fig:Dispersion} we have provided the dispersion diagrams of the tunable CRLH unit cell for different varactor values based on the simulation in HFSS along with the air line which determines the fast wave region, i.e., the frequency region below the air line. Considering the applied bias voltages to the varactors, solid red line shows the dispersion diagram under ``Bias 0" state which can provide $\beta p=-18^\circ$ around $1.95$ GHz and the solid blue line is the dispersion diagram of the unit cell under the "Bias 1" state which can provide $\beta p=15^\circ$ around $1.95$ GHz, as stated in the manuscript. Furthermore, as shown in the figure, a bandwidth of $120$ MHz around $1.95$  GHz is still in the fast wave region and $\beta p$ is positive under the ``Bias 1" and negative under the ``Bias 0". We note that in order to increase the bandwidth, a unit cell with a more slower slope dispersion diagram may be designed to expand the fast wave frequency region if needed.
f

\newpage
\bibliographystyle{IEEEtran}

\begin{thebibliography}{10}
\providecommand{\url}[1]{#1}
\csname url@samestyle\endcsname
\providecommand{\newblock}{\relax}
\providecommand{\bibinfo}[2]{#2}
\providecommand{\BIBentrySTDinterwordspacing}{\spaceskip=0pt\relax}
\providecommand{\BIBentryALTinterwordstretchfactor}{4}
\providecommand{\BIBentryALTinterwordspacing}{\spaceskip=\fontdimen2\font plus
\BIBentryALTinterwordstretchfactor\fontdimen3\font minus
  \fontdimen4\font\relax}
\providecommand{\BIBforeignlanguage}[2]{{%
\expandafter\ifx\csname l@#1\endcsname\relax
\typeout{** WARNING: IEEEtran.bst: No hyphenation pattern has been}%
\typeout{** loaded for the language `#1'. Using the pattern for}%
\typeout{** the default language instead.}%
\else
\language=\csname l@#1\endcsname
\fi
#2}}
\providecommand{\BIBdecl}{\relax}
\BIBdecl

\bibitem{Ning_survey_2021}
N.~Xie, Z.~Li, and H.~Tan, ``A survey of physical-layer authentication in
  wireless communications,'' \emph{IEEE Communications Surveys \& Tutorials},
  vol.~23, no.~1, pp. 282--310, 2021.

\bibitem{Sina_Alireza_survey2019}
S.~Rezaei~Aghdam, A.~Nooraiepour, and T.~M. Duman, ``An overview of physical
  layer security with finite-alphabet signaling,'' \emph{IEEE Communications
  Surveys \& Tutorials}, vol.~21, no.~2, pp. 1829--1850, 2019.

\bibitem{Hu_2019}
X.~Hu, C.~Liu, S.~Liu, W.~You, Y.~Li, and Y.~Zhao, ``A systematic analysis
  method for 5g non-access stratum signalling security,'' \emph{IEEE Access},
  vol.~7, pp. 125\,424--125\,441, 2019.

\bibitem{nooraiepour_TCCN}
A.~Nooraiepour, W.~U. Bajwa, and N.~B. Mandayam, ``Learning-aided physical
  layer attacks against multicarrier communications in iot,'' \emph{IEEE
  Transactions on Cognitive Communications and Networking}, vol.~7, no.~1, pp.
  239--254, 2021.

\bibitem{nooraiepour_MILCOM}
A.~Nooraiepour, K.~Hamidouche, W.~U. Bajwa, and N.~Mandayam, ``How secure are
  multicarrier communication systems against signal exploitation attacks?'' in
  \emph{MILCOM 2018 - 2018 IEEE Military Communications Conference (MILCOM)},
  2018, pp. 201--206.

\bibitem{ding2013vector}
Y.~Ding and V.~F. Fusco, ``A vector approach for the analysis and synthesis of
  directional modulation transmitters,'' \emph{IEEE Transactions on Antennas
  and Propagation}, vol.~62, no.~1, pp. 361--370, 2013.

\bibitem{daly2009directional}
M.~P. Daly and J.~T. Bernhard, ``Directional modulation technique for phased
  arrays,'' \emph{IEEE Transactions on Antennas and Propagation}, vol.~57,
  no.~9, pp. 2633--2640, 2009.

\bibitem{OFDM_DM}
Y.~Ding, V.~Fusco, J.~Zhang, and W.-Q. Wang, ``Time-modulated ofdm directional
  modulation transmitters,'' \emph{IEEE Transactions on Vehicular Technology},
  vol.~68, no.~8, pp. 8249--8253, 2019.

\bibitem{OurACM}
\BIBentryALTinterwordspacing
A.~Nooraiepour, S.~Vosoughitabar, C.-T.~M. Wu, W.~U. Bajwa, and N.~B. Mandayam,
  ``Time-varying metamaterial-enabled directional modulation schemes for
  physical layer security in wireless communication links,'' \emph{J. Emerg.
  Technol. Comput. Syst.}, vol.~18, no.~4, oct 2022. [Online]. Available:
  \url{https://doi.org/10.1145/3513088}
\BIBentrySTDinterwordspacing

\bibitem{OurIMS}
S.~Vosoughitabar, A.~Nooraiepour, W.~U. Bajwa, N.~Mandayam, and C.-T.~M. Wu,
  ``Metamaterial-enabled 2d directional modulation array transmitter for
  physical layer security in wireless communication links,'' in \emph{2022
  IEEE/MTT-S International Microwave Symposium - IMS 2022}, 2022, pp. 595--598.

\bibitem{nature_elec_1}
S.~Venkatesh, X.~Lu, B.~Tang, and K.~Sengupta, ``Secure space--time-modulated
  millimetre-wave wireless links that are resilient to distributed eavesdropper
  attacks,'' \emph{Nature Electronics}, vol.~4, no.~11, pp. 827--836, 2021.

\bibitem{nature_elec_2}
S.~Venkatesh, H.~Saeidi, K.~Sengupta, and X.~Lu, ``Millimeter-wave physical
  layer security through space-time modulated transmitter arrays,'' in
  \emph{2022 IEEE 22nd Annual Wireless and Microwave Technology Conference
  (WAMICON)}, 2022, pp. 1--4.

\bibitem{daly2010demonstration}
M.~P. Daly, E.~L. Daly, and J.~T. Bernhard, ``Demonstration of directional
  modulation using a phased array,'' \emph{IEEE Transactions on Antennas and
  Propagation}, vol.~58, no.~5, pp. 1545--1550, 2010.

\bibitem{mendez2016hybrid}
R.~M{\'e}ndez-Rial, C.~Rusu, N.~Gonz{\'a}lez-Prelcic, A.~Alkhateeb, and R.~W.
  Heath, ``Hybrid {MIMO} architectures for millimeter wave communications:
  Phase shifters or switches?'' \emph{IEEE Access}, vol.~4, pp. 247--267, 2016.

\bibitem{yu2014flat}
N.~Yu and F.~Capasso, ``Flat optics with designer metasurfaces,'' \emph{Nature
  materials}, vol.~13, no.~2, pp. 139--150, 2014.

\bibitem{chen2016review}
H.-T. Chen, A.~J. Taylor, and N.~Yu, ``A review of metasurfaces: physics and
  applications,'' \emph{Reports on progress in physics}, vol.~79, no.~7, p.
  076401, 2016.

\bibitem{glybovski2016metasurfaces}
S.~B. Glybovski, S.~A. Tretyakov, P.~A. Belov, Y.~S. Kivshar, and C.~R.
  Simovski, ``Metasurfaces: From microwaves to visible,'' \emph{Physics
  reports}, vol. 634, pp. 1--72, 2016.

\bibitem{Shaghayegh_IMS2020}
S.~Vosoughitabar, M.~Zhu, and C.-T.~M. Wu, ``A distributed mixer-based
  nonreciprocal {CRLH} leaky wave antenna for simultaneous transmit and
  receive,'' in \emph{2021 IEEE MTT-S International Microwave Symposium (IMS)},
  2021, pp. 408--411.

\bibitem{R2-A-wireless-communication-scheme}
\BIBentryALTinterwordspacing
L.~Zhang, M.~Z. Chen, W.~Tang, J.~Y. Dai, L.~Miao, X.~Y. Zhou, S.~Jin,
  Q.~Cheng, and T.~J. Cui, ``A wireless communication scheme based on space-
  and frequency-division multiplexing using digital metasurfaces,''
  \emph{Nature Electronics}, vol.~4, no.~3, pp. 218--227, Mar 2021. [Online].
  Available: \url{https://doi.org/10.1038/s41928-021-00554-4}
\BIBentrySTDinterwordspacing

\bibitem{R2-Breaking-Reciprocity}
\BIBentryALTinterwordspacing
L.~Zhang, X.~Q. Chen, R.~W. Shao, J.~Y. Dai, Q.~Cheng, G.~Castaldi, V.~Galdi,
  and T.~J. Cui, ``Breaking reciprocity with space-time-coding digital
  metasurfaces,'' \emph{Advanced Materials}, vol.~31, no.~41, p. 1904069, 2019.
  [Online]. Available:
  \url{https://onlinelibrary.wiley.com/doi/abs/10.1002/adma.201904069}
\BIBentrySTDinterwordspacing

\bibitem{cui2014coding}
T.~J. Cui, M.~Q. Qi, X.~Wan, J.~Zhao, and Q.~Cheng, ``Coding metamaterials,
  digital metamaterials and programmable metamaterials,'' \emph{Light: Science
  \& Applications}, vol.~3, no.~10, pp. e218--e218, 2014.

\bibitem{li2017electromagnetic}
L.~Li, T.~J. Cui, W.~Ji, S.~Liu, J.~Ding, X.~Wan, Y.~B. Li, M.~Jiang, C.-W.
  Qiu, and S.~Zhang, ``Electromagnetic reprogrammable coding-metasurface
  holograms,'' \emph{Nature communications}, vol.~8, no.~1, pp. 1--7, 2017.

\bibitem{zhang2018space}
L.~Zhang, X.~Q. Chen, S.~Liu, Q.~Zhang, J.~Zhao, J.~Y. Dai, G.~D. Bai, X.~Wan,
  Q.~Cheng, G.~Castaldi, V.~Galdi, and T.~J. Cui, ``Space-time-coding digital
  metasurfaces,'' \emph{Nature communications}, vol.~9, no.~1, pp. 1--11, 2018.

\bibitem{liu2016anisotropic}
S.~Liu, T.~J. Cui, Q.~Xu, D.~Bao, L.~Du, X.~Wan, W.~X. Tang, C.~Ouyang, X.~Y.
  Zhou, H.~Yuan, H.~F. Ma, W.~X. Jiang, J.~Han, W.~Zhang, and Q.~Cheng,
  ``Anisotropic coding metamaterials and their powerful manipulation of
  differently polarized terahertz waves,'' \emph{Light: Science \&
  Applications}, vol.~5, no.~5, pp. e16\,076--e16\,076, 2016.

\bibitem{R2-Programmable-time-domain}
\BIBentryALTinterwordspacing
J.~Zhao, X.~Yang, J.~Y. Dai, Q.~Cheng, X.~Li, N.~H. Qi, J.~C. Ke, G.~D. Bai,
  S.~Liu, S.~Jin, A.~Alù, and T.~J. Cui, ``{Programmable time-domain
  digital-coding metasurface for non-linear harmonic manipulation and new
  wireless communication systems},'' \emph{National Science Review}, vol.~6,
  no.~2, pp. 231--238, 11 2018. [Online]. Available:
  \url{https://doi.org/10.1093/nsr/nwy135}
\BIBentrySTDinterwordspacing

\bibitem{R2-Realization-of-Multi-Modulation}
J.~Y. Dai, W.~Tang, L.~X. Yang, X.~Li, M.~Z. Chen, J.~C. Ke, Q.~Cheng, S.~Jin,
  and T.~J. Cui, ``Realization of multi-modulation schemes for wireless
  communication by time-domain digital coding metasurface,'' \emph{IEEE
  Transactions on Antennas and Propagation}, vol.~68, no.~3, pp. 1618--1627,
  2020.

\bibitem{basar2019wireless}
E.~Basar, M.~Di~Renzo, J.~De~Rosny, M.~Debbah, M.-S. Alouini, and R.~Zhang,
  ``Wireless communications through reconfigurable intelligent surfaces,''
  \emph{IEEE Access}, vol.~7, pp. 116\,753--116\,773, 2019.

\bibitem{wu2019intelligent}
Q.~Wu and R.~Zhang, ``Intelligent reflecting surface enhanced wireless network
  via joint active and passive beamforming,'' \emph{IEEE Transactions on
  Wireless Communications}, vol.~18, no.~11, pp. 5394--5409, 2019.

\bibitem{wu2019towards}
------, ``Towards smart and reconfigurable environment: Intelligent reflecting
  surface aided wireless network,'' \emph{IEEE Communications Magazine},
  vol.~58, no.~1, pp. 106--112, 2019.

\bibitem{Perfect_Absorption_Meta_Atom}
\BIBentryALTinterwordspacing
M.~F.~Imani, D.~R. Smith, and P.~del Hougne, ``Perfect absorption in a
  disordered medium with programmable meta-atom inclusions,'' \emph{Advanced
  Functional Materials}, vol.~30, no.~52, p. 2005310, 2020. [Online].
  Available:
  \url{https://onlinelibrary.wiley.com/doi/abs/10.1002/adfm.202005310}
\BIBentrySTDinterwordspacing

\bibitem{MassiveBackscatter}
\BIBentryALTinterwordspacing
H.~Zhao, Y.~Shuang, M.~Wei, T.~J. Cui, P.~d. Hougne, and L.~Li,
  ``Metasurface-assisted massive backscatter wireless communication with
  commodity wi-fi signals,'' \emph{Nature Communications}, vol.~11, no.~1, p.
  3926, Aug 2020. [Online]. Available:
  \url{https://doi.org/10.1038/s41467-020-17808-y}
\BIBentrySTDinterwordspacing

\bibitem{caloz2005electromagnetic}
C.~Caloz and T.~Itoh, \emph{Electromagnetic metamaterials: transmission line
  theory and microwave applications}.\hskip 1em plus 0.5em minus 0.4em\relax
  John Wiley \& Sons, 2005.

\bibitem{caloz2008crlh}
C.~Caloz, T.~Itoh, and A.~Rennings, ``{CRLH} metamaterial leaky-wave and
  resonant antennas,'' \emph{IEEE Antennas and Propagation Magazine}, vol.~50,
  no.~5, pp. 25--39, 2008.

\bibitem{jackson2012leaky}
D.~R. Jackson, C.~Caloz, and T.~Itoh, ``Leaky-wave antennas,''
  \emph{Proceedings of the IEEE}, vol. 100, no.~7, pp. 2194--2206, 2012.

\bibitem{yuan2019multi}
Y.~Yuan, C.~Lu, A.~Y.-K. Chen, C.-H. Tseng, and C.-T.~M. Wu, ``Multi-target
  concurrent vital sign and location detection using metamaterial-integrated
  self-injection-locked quadrature radar sensor,'' \emph{IEEE Transactions on
  Microwave Theory and Techniques}, vol.~67, no.~12, pp. 5429--5437, 2019.

\bibitem{salarkaleji2016two}
M.~Salarkaleji, M.~A. Ali, and C.-T.~M. Wu, ``Two-dimensional full-hemisphere
  frequency scanning array based on metamaterial leaky wave antennas and feed
  networks,'' in \emph{2016 IEEE MTT-S International Microwave Symposium
  (IMS)}.\hskip 1em plus 0.5em minus 0.4em\relax IEEE, 2016, pp. 1--4.

\bibitem{lim2004metamaterial}
S.~Lim, C.~Caloz, and T.~Itoh, ``Metamaterial-based electronically controlled
  transmission-line structure as a novel leaky-wave antenna with tunable
  radiation angle and beamwidth,'' \emph{IEEE Transactions on Microwave Theory
  and Techniques}, vol.~52, no.~12, pp. 2678--2690, 2004.

\bibitem{gil2006tunable}
I.~Gil, J.~Bonache, J.~Garcia-Garcia, and F.~Martin, ``Tunable metamaterial
  transmission lines based on varactor-loaded split-ring resonators,''
  \emph{IEEE transactions on Microwave theory and techniques}, vol.~54, no.~6,
  pp. 2665--2674, 2006.

\bibitem{luo2020active}
Y.~Luo, K.~Qin, H.~Ke, B.~Xu, S.~Xu, and G.~Yang, ``Active metamaterial antenna
  with beam scanning manipulation based on a digitally modulated array factor
  method,'' \emph{IEEE Transactions on Antennas and Propagation}, vol.~69,
  no.~2, pp. 1198--1203, 2020.

\bibitem{shlezinger2019dynamic}
N.~Shlezinger, O.~Dicker, Y.~C. Eldar, I.~Yoo, M.~F. Imani, and D.~R. Smith,
  ``Dynamic metasurface antennas for uplink massive mimo systems,'' \emph{IEEE
  Transactions on Communications}, vol.~67, no.~10, pp. 6829--6843, 2019.

\bibitem{boyarsky2021electronically}
M.~Boyarsky, T.~Sleasman, M.~F. Imani, J.~N. Gollub, and D.~R. Smith,
  ``Electronically steered metasurface antenna,'' \emph{Scientific reports},
  vol.~11, no.~1, pp. 1--10, 2021.

\bibitem{R2-Sideband-free-space}
\BIBentryALTinterwordspacing
G.-B. Wu, J.~Y. Dai, Q.~Cheng, T.~J. Cui, and C.~H. Chan, ``Sideband-free
  space--time-coding metasurface antennas,'' \emph{Nature Electronics}, vol.~5,
  no.~11, pp. 808--819, Nov 2022. [Online]. Available:
  \url{https://doi.org/10.1038/s41928-022-00857-0}
\BIBentrySTDinterwordspacing

\bibitem{vosoughitabar2023programming}
S.~Vosoughitabar and C.-T.~M. Wu, ``Programming nonreciprocity and harmonic
  beam steering via a digitally space-time-coded metamaterial antenna,''
  \emph{Scientific Reports}, vol.~13, no.~1, p. 7338, 2023.

\bibitem{Wyner}
A.~D. Wyner, ``The wire-tap channel,'' \emph{The Bell System Technical
  Journal}, vol.~54, no.~8, pp. 1355--1387, 1975.

\bibitem{B&B2}
M.~Lee, G.~Yu, and G.~Y. Li, ``Learning to branch: Accelerating resource
  allocation in wireless networks,'' \emph{IEEE Transactions on Vehicular
  Technology}, vol.~69, no.~1, pp. 958--970, 2020.

\bibitem{BranchandBound}
\BIBentryALTinterwordspacing
A.~H. Land and A.~G. Doig, ``An automatic method of solving discrete
  programming problems,'' \emph{Econometrica}, vol.~28, no.~3, pp. 497--520,
  1960. [Online]. Available: \url{http://www.jstor.org/stable/1910129}
\BIBentrySTDinterwordspacing

\bibitem{Sina_MIMO}
S.~R. Aghdam and T.~M. Duman, ``Joint precoder and artificial noise design for
  mimo wiretap channels with finite-alphabet inputs based on the cut-off
  rate,'' \emph{IEEE Transactions on Wireless Communications}, vol.~16, no.~6,
  pp. 3913--3923, 2017.

\bibitem{Channel_Estimation}
S.~Coleri, M.~Ergen, A.~Puri, and A.~Bahai, ``Channel estimation techniques
  based on pilot arrangement in ofdm systems,'' \emph{IEEE Transactions on
  Broadcasting}, vol.~48, no.~3, pp. 223--229, 2002.

\bibitem{B&B4}
\BIBentryALTinterwordspacing
H.~He, H.~Daume~III, and J.~M. Eisner, ``Learning to search in branch and bound
  algorithms,'' in \emph{Advances in Neural Information Processing Systems},
  Z.~Ghahramani, M.~Welling, C.~Cortes, N.~Lawrence, and K.~Weinberger, Eds.,
  vol.~27.\hskip 1em plus 0.5em minus 0.4em\relax Curran Associates, Inc.,
  2014. [Online]. Available:
  \url{https://proceedings.neurips.cc/paper/2014/file/757f843a169cc678064d9530d12a1881-Paper.pdf}
\BIBentrySTDinterwordspacing

\bibitem{meroune}
A.~Zappone, M.~D. Renzo, and M.~Debbah, ``Wireless networks design in the era
  of deep learning: {Model-Based}, {AI-Based}, or {Both}?'' \emph{IEEE
  Transactions on Communications}, vol.~67, pp. 7331--7376, 2019.

\bibitem{HyPhyLearn}
A.~Nooraiepour, W.~U. Bajwa, and N.~B. Mandayam, ``A hybrid model-based and
  learning-based approach for classification using limited number of training
  samples,'' \emph{IEEE Open Journal of Signal Processing}, vol.~3, pp. 49--70,
  2022.

\bibitem{MLSP}
------, ``Hyphylearn: A domain adaptation-inspired approach to classification
  using limited number of training samples,'' in \emph{2021 IEEE 31st
  International Workshop on Machine Learning for Signal Processing (MLSP)},
  2021, pp. 1--6.

\bibitem{1381686}
S.~Lim, C.~Caloz, and T.~Itoh, ``Metamaterial-based electronically controlled
  transmission-line structure as a novel leaky-wave antenna with tunable
  radiation angle and beamwidth,'' \emph{IEEE Transactions on Microwave Theory
  and Techniques}, vol.~53, no.~1, pp. 161--173, 2005.

\end{thebibliography}

\end{document}